\documentclass[a4paper,12pt]{article}

 \usepackage[dvips]{graphicx}
 \usepackage{color}
 \usepackage{ulem}

\author{Iwata and Umeno}
\title{Correlation Analysis for Total Electron Content Anomalies on 11th March, 2011}
\date{}

\begin{document}
\maketitle
Takuya Iwata,
Ken Umeno\\

Department of Applied Mathematics and Physics, Graduate school of Informatics, Kyoto University, Yoshidahon-machi, Sakyo-ku, Kyoto, Japan

\section{Introduction}
Ionosphere is a shell of a large amount of electrons and it is disturbed by various causes such as volcanic eruptions \cite{heki2006,Dautermann2009}, solar flares \cite{Donnelly1976}, earthquakes \cite{Cahyadi2015}, and so on.
Analyzing Total Electron Content (TEC) data is one of the most popular method to monitor the ionosphere.
TEC data contains the number of electrons integrated along the line of sight between the GNSS satellites and the observation stations on the ground.
Japan has a dense GNSS observation network (the GNSS earth observation network, GEONET) to collect daily TEC data in Japan.
GEONET is composed of more than 1000 stations and collects TEC data everyday.
The GNSS data from GEONET are available freely for everyone.\\

Ionospheric disturbances shortly after large earthquakes (Coseismic ionospheric disturbances, CID) have been observed and reported for several cases \cite{Calais1995,Ducic2003,Liu2010}.
As for the 2011 Tohoku-Oki earthquake, various kinds of ionospheric disturbances following the earthquake have been studied \cite{Astafyeva2011,Tsugawa2011}.\\

Preseismic ionospheric disturbances have also been studied and some signs before large earthquakes have been reported \cite{Heki2011,Heki2015,Jin2015}.
Thus, it is important to clarify whether such preseismic anomalies really exist or not and establish a valid analysis method to predict large earthquakes if preseismic anomalies exist.
Such analysis methods to detect preseismic ionospheric anomalies with the use of TEC data have not been established until now.
In this paper, we propose the correlation analysis method to detect TEC anomalies.
This method is practical because it does not need data after the corresponding earthquakes and can be easily implemented as automatic computation of correlations.
\section{TEC data}
TEC changes over time are calculated by analyzing the phase differences between the two carrier waves with the different frequencies (1.2 GHz and 1.5 GHz) from GNSS satellites.
TEC values obtained by analysing GNSS signals are largely affected by the elevation angle of the satellites, i.e. the lower elevation angle increasing the apparent penetration length of the line of sight, and results in larger slant TEC values.
That is, TEC value is susceptible to not only the condition of the ionosphere but also to the elevation angle of the satellites.
This fact is one of the major obstacles to TEC data analysis.
To overcome this difficulty, vertical TEC (VTEC) is used.
We can get VTEC data by computing the vertical components of TEC data, i.e. multiplying TEC by the cosine of the elevation angle of the satellites and we can get VTEC.\\

Not only TEC changes over time, but also the positions of GNSS satellites at each time are importnat information.
Following the custom for simplification, we make an assumption that there is a thin layer about 300 kilometers above us and calculate the intersection of the line of sight with this layer.
Such an intersection is called as Ionospheric Pierce Point (IPP), and its projection onto the ground is called as Sub-ionospheric Point (SIP).
IPP and SIP give the approximate position where the TEC data are derived.\\

GEONET has a huge amount of TEC data in Japan, but only limited amount of data are freely available via the Internet, i.e. TEC data for the last three or four years are freely available.
Here, we used TEC data obtained before the 2011 Tohoku-Oki earthquake (Mw 9.0), which is the largest earthquake in Japan during this period.
 We used TEC data from 2 months before the earthquake to the earthquake day for checking the validity of our proposed method to be described in the next section.

\section{Analysis method}
\subsection{correlation analysis}
The analysis method we propose here is based on the idea of applications of the correlation detection method used in VLBI (Very Long Baseline Interferometry) and spreading spectrum communications technology to the analysis of TEC anomalies and given as follows:\\
STEP 0. Choose the central GNSS station and set up three parameters, i.e. $t_{sample}$, $t_{test}$, and $M$.
$t_{sample}$ is the length of data used for regression \textit{training} and $t_{test}$ is the length of data used for regression \textit{test}, and the value $M$ denotes the number of GNSS stations we use.\\
STEP 1. At each station $i$ and at each time epoch $t$, let $Sample Data$ be the data from $t$ to $t+t_{sample}$ and $Test Data$ be the data from $t+t_{sample}$ to $t+t_{sample}+t_{test}$.\\
STEP 2. Fit a curve to $Sample Data$ by the least square method.\\
STEP 3. Calculate a deviation of the $Test Data$ from the model curve, representing as an "anomaly".  The anomaly at station $i$ at time $t^\prime$ is denoted as $x_{i, t^\prime}$, where $-\infty < x_{i, t^\prime} < \infty$ and we assume $<x_{i, t^{\prime}}> = 0$.\\
STEP 4. Calculate a summation of correlations between the anomalies at the central GNSS station and the surrounding stations as follows:\\
\begin{equation}
C(T) = \frac{1}{N\times M}\sum_{i=1}^{M}\sum_{j=0}^{N-1}x_{i, t+t_{sample}+j\Delta t}x_{0,t+t_{sample}+j\Delta t}
\end{equation}
\[
T=t+t_{sample}+t_{test}
\]
Here, $N$ is the number of data in \textit{TestData}, $\Delta t$ is a sampling interval in \textit{TestData}, which means $\Delta t = t_{test}/(N-1)$, and $i=0$ means the central GNSS station, where $t_{sample} = 2.0$ [hours], $t_{test} = 0.25$ [hours] and $M=30$ [stations].\\
$t_{sample}$ and $t_{test}$ are significant parameters.
If $t_{sample}$ is too large, we need a lot of data to calculate $C$($T$) and if $t_{sample}$ is too small, we can not obtain reliable model curves in STEP 2.
If $t_{test}$ is too large, deviations of the $TestData$ from the model curves become large and if $t_{test}$ is too small, we can not catch the changes of TEC data accurately.
For these reasons, we set these parameters as written above.
In STEP 1, if there are insufficient number of $Sample Data$ or $Test Data$, then let $C$($T$) be NA (missing value).\\
In STEP 2, we can use arbitrary functions such as polynomial functions, Fourier series, and Gaussian functions for nonlinear regression.
In this paper, we use a 7th polynomial function, Fourier series and Gaussian functions for data extrapolation.
In later section, we investigate the effectiveness of using such functions for nonlinear regression.\\
The results do not depend on our choice of extrapolating functions, but they sensitively depend on the parameters set up in STEP 0.

\subsection{Fitting to TEC time series}
Generally, fitting a curve to data, i.e. nonlinear regression, needs to consider some points.\\
As a result of regression analysis, we get a function $f(t)$ as a model function for TEC data.
Theoretically, we can choose arbitrary functions for $f(t)$.\\
For example, the most simple and often-used functions are polynomial functions.
A $D$-th polynomial function is defined by
\begin{equation}
f(t) = \sum_{i=0}^Da_it^i.
\end{equation}
The number of parameters is $D+1$.
Such polynomial functions are simple but not the best candidate since $f(t)$ increase or decrease infinitely as $t$ increases whereas the data do not.\\
On the contrary, Fourier series are periodic and bounded functions.
A $D$-th Fourier series is defined by
\begin{equation}
f(t) = a_0 + \sum_{k=1}^D\left\{a_k\sin \left(\frac{2\pi k}{T}t\right)+b_k\cos \left(\frac{2\pi k}{T}t\right)\right\}.
\end{equation}
The number of parameters is $2D+1$ and $T$ means the period of data.
In this case, setting $T=24$ [hours] is good since GNSS satellites have a period of approximately 24 hours.\\
We can also use Gaussian functions for $f$($t$).
A $D$-th order Gaussian extrapolating function is defined by
\begin{equation}
f(t) = a_0 + \sum_{i=1}^Da_i{\rm exp}\left(-\frac{(t-\mu_i)^2}{2\sigma_i^2}\right).
\end{equation}
The number of parameters is $D+1$.
Hyper parameters $\mu_i$ and $\sigma_i$ have to be set in advance.

\newpage
\section{Results}
\subsection{TEC anomaly near the epicenter on 11th March, 2011}
Figure \ref{kitaibaraki70} shows the result of correlation analysis on 11th March, 2011 when the 2011 Tohoku-Oki earthquake occured.
We chose the GPS satellite 26 and the 0214 (Kitaibaraki) GNSS station as the central station.
The x-axis is time $T$ in the coodinated universal time UTC and the y-axis represents an accumulated correlation, here, briefly wirriten $C$($T$) defined in Eq. (1).
The black line represents T=05:46 [UT] , i.e., the \textit{exact} time when the 2011 Tohoku-Oki earthquake occured.
Here, we chose a 7th polynomial function as a refference curve.\\
This result indicates that there \textit{exists} an anomalous trend \textit{before} the earthquake.\\
Figure \ref{sip} shows the tracks of SIP and the 50 GNSS stations surrounding the Kitaibaraki (0214) station used for the correlation analysis to get Fig.1.
\subsection{TEC observation on non-earthquake days}
Figure \ref{kitaibaraki30-60} shows the results of correlation analysis on non-earthquake days such as 2011/02/19 and 2011/03/01.
The central GNSS station used for the analysis is the same 0214 station (Kitaibaraki).
In comparison to Fig. \ref{kitaibaraki70}, correlations $C$($T$) are quite small and actually quiet.
The maximal value of the absolute value of correlation $C$($T$) on these days is at most 5, whereas $C$($T$) just before the earthquake on March 11, 2011 recorded more than 25, which is five times than the maximal value of absolute value of correlation of these \textit{normal} days.

Figure \ref{tec-cor} compares the real TEC data and the results of our correlation analysis.
Correlation analysis detected the anomaly that is difficult to detect by merely looking real TEC data.\\

\subsection{TEC observation all over Japan on 11th March, 2011}
Figure \ref{japan70} shows the change of $C$($T$) at all stations in Japan on the earthquake day, 11 th March, 2011.
These results suggest that the anomalies can be seen near the epicenter and just on the day of the earthquake occured.
In Fig. \ref{japan70}, the anomalies can be seen also in southwest Japan.
This anomalous area is smaller than near the epicenter.
Although we are not sure of the origin of this anomalous area, it could be explanation of this phenomenon that these southwest Japan area is near the plate boundary as shown in Fig. \ref{bound}.

\begin{figure}
\noindent\includegraphics[width=35pc]{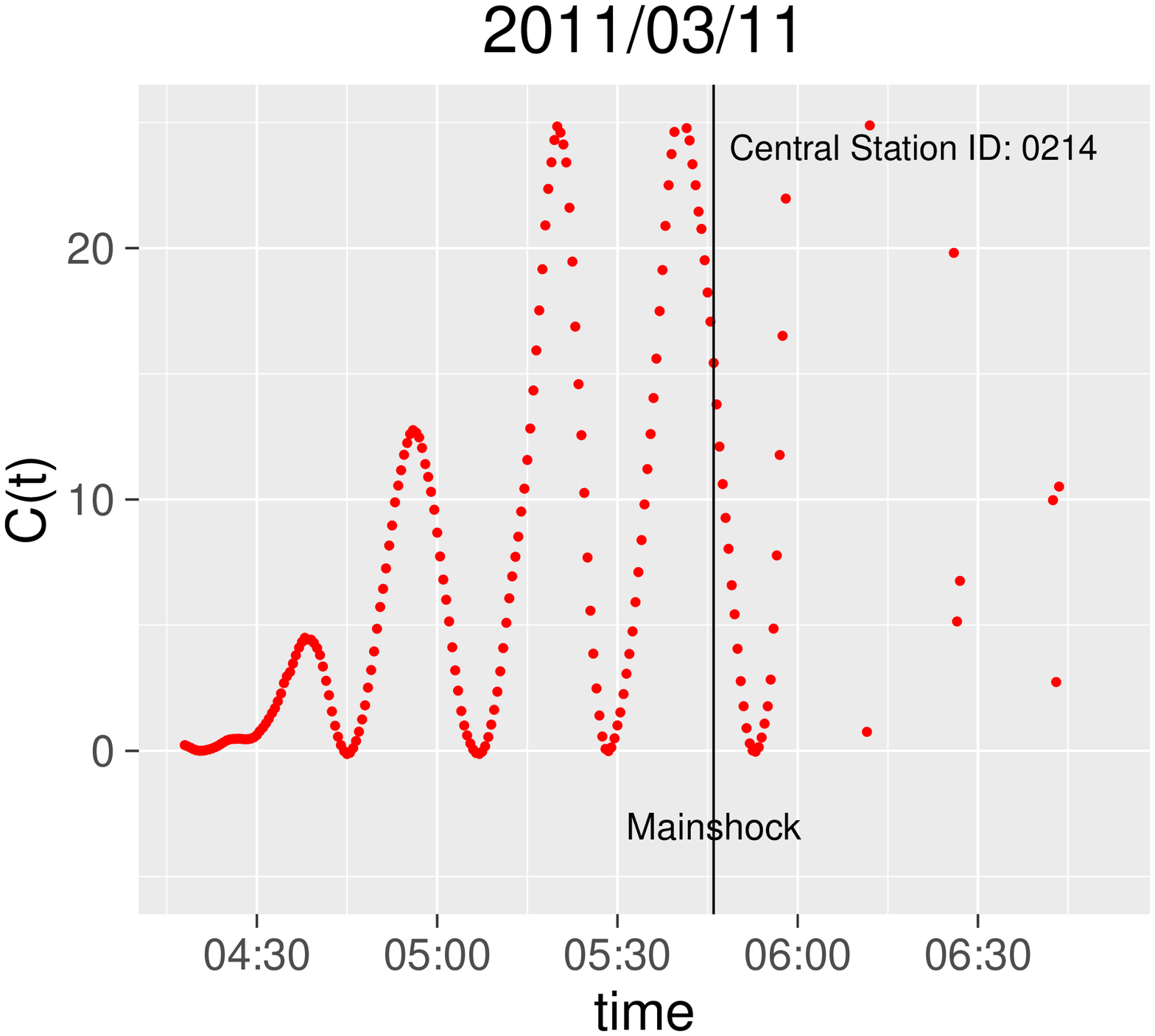}
\caption{The result of correlation analysis on 11th March, 2011.  The vertical axis shows the correlation $C(T)$ and the horizontal one the time $t$ [UTC].  The black line indicates the exact time 05:46 [UTC] when the 2011 Tohoku-Oki earthquake occured.}
\label{kitaibaraki70}
\end{figure}

\begin{figure}
\noindent\includegraphics[width=30pc]{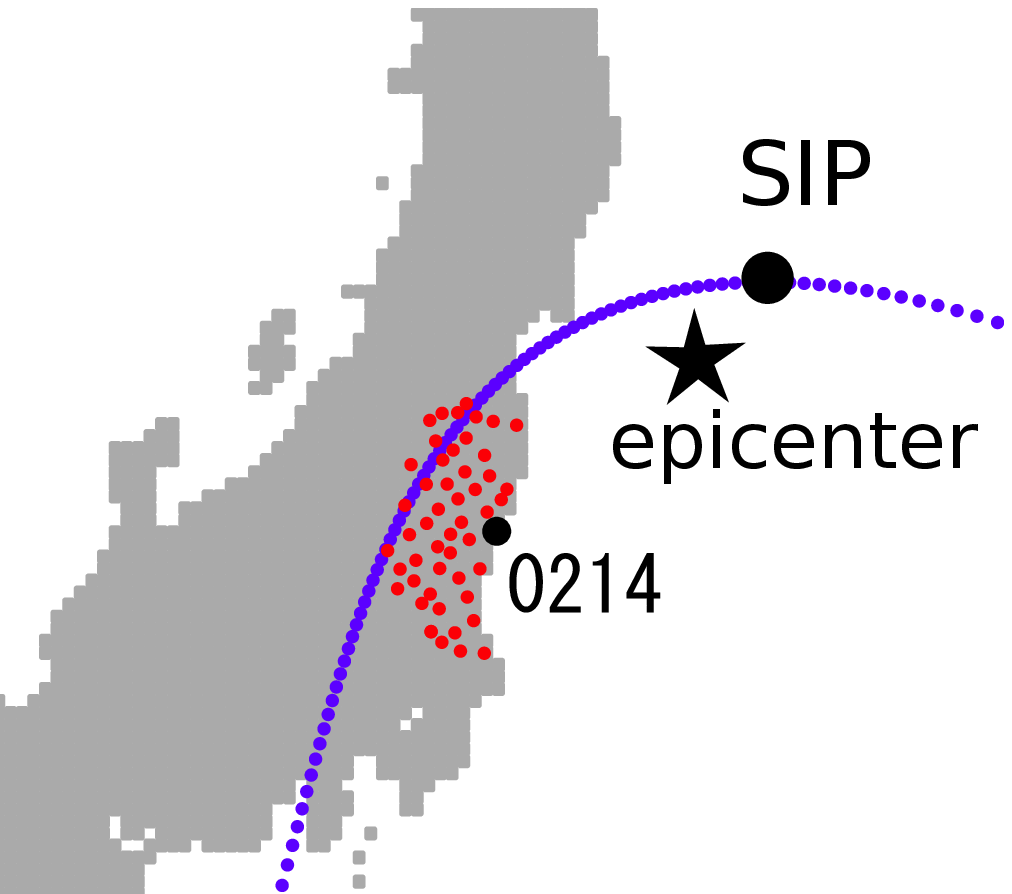}
\caption{The blue line represents the SIP track of the pair of the 0214 station (Kitaibaraki) and the GPS satellite 26, and the red points represent the location of surrounding 50 stations around the central station.  The star represents the epicenter of the 2011 Tohoku-Oki earthquake, and the black circle on the blue line represents the SIP position at the earthquake occurrence time.}
\label{sip}
\end{figure}

\begin{figure}
\begin{tabular}{cc}
\includegraphics[height=16pc]{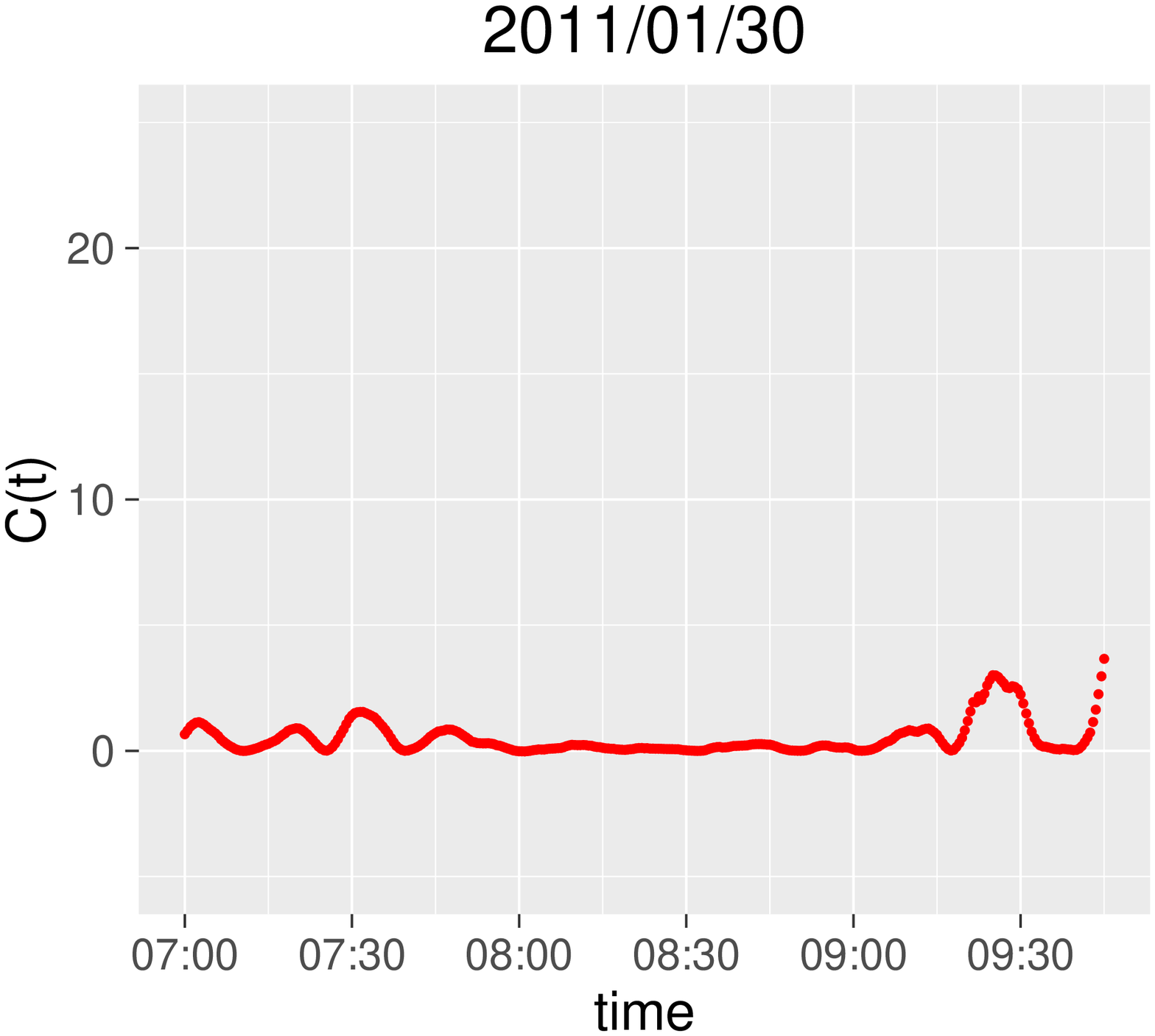} &
\includegraphics[height=16pc]{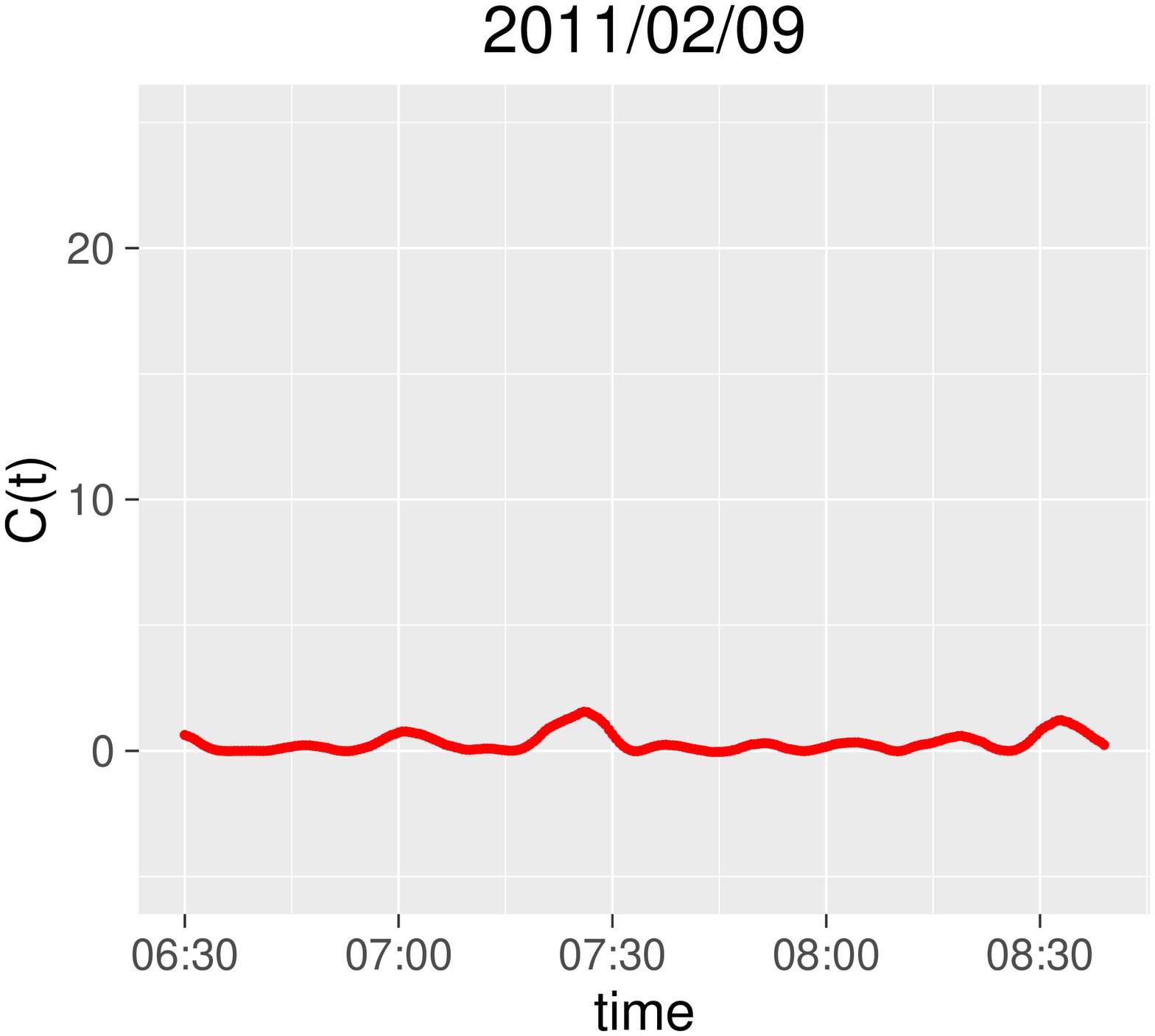}\\
\includegraphics[height=16pc]{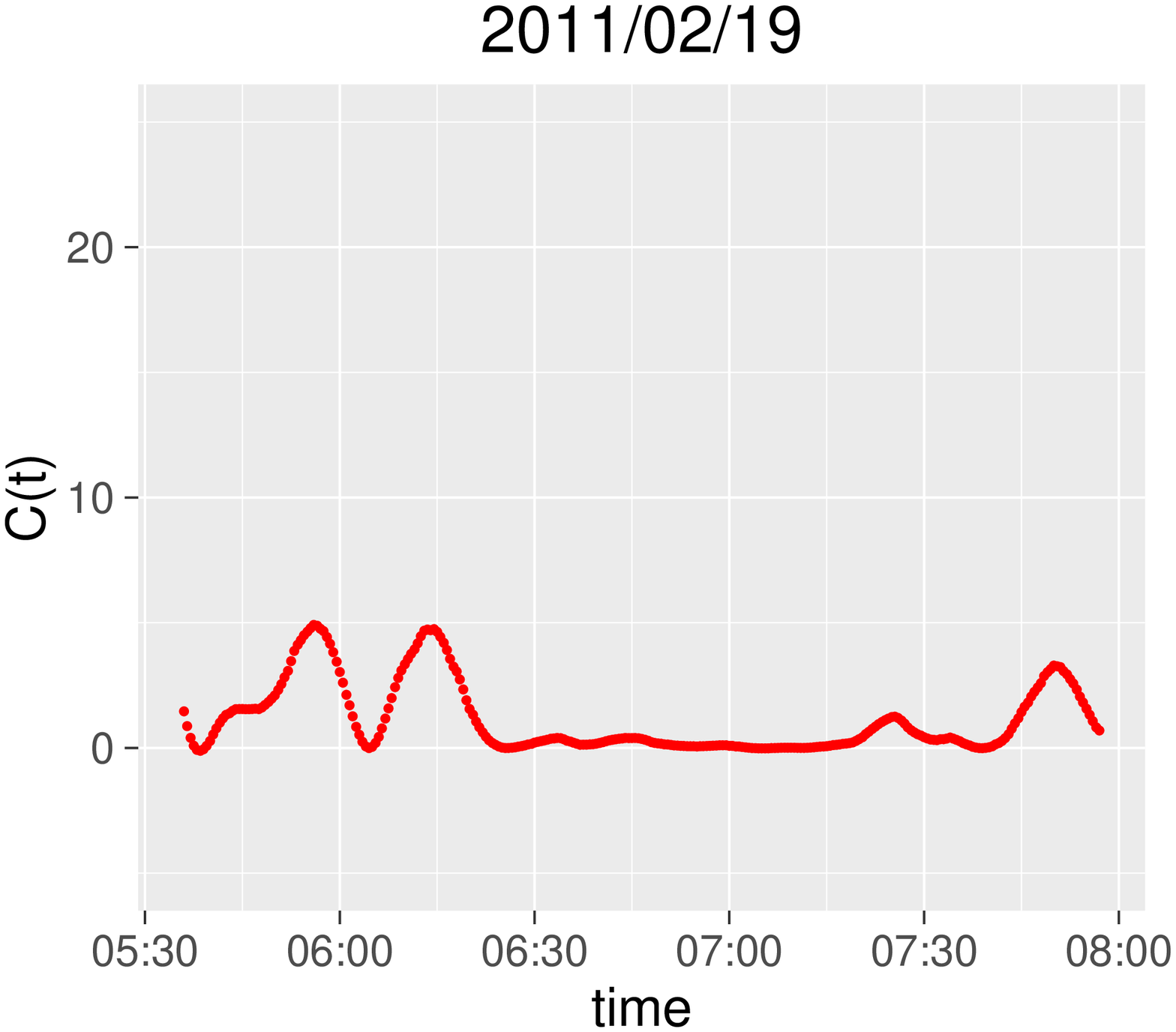} &
\includegraphics[height=16pc]{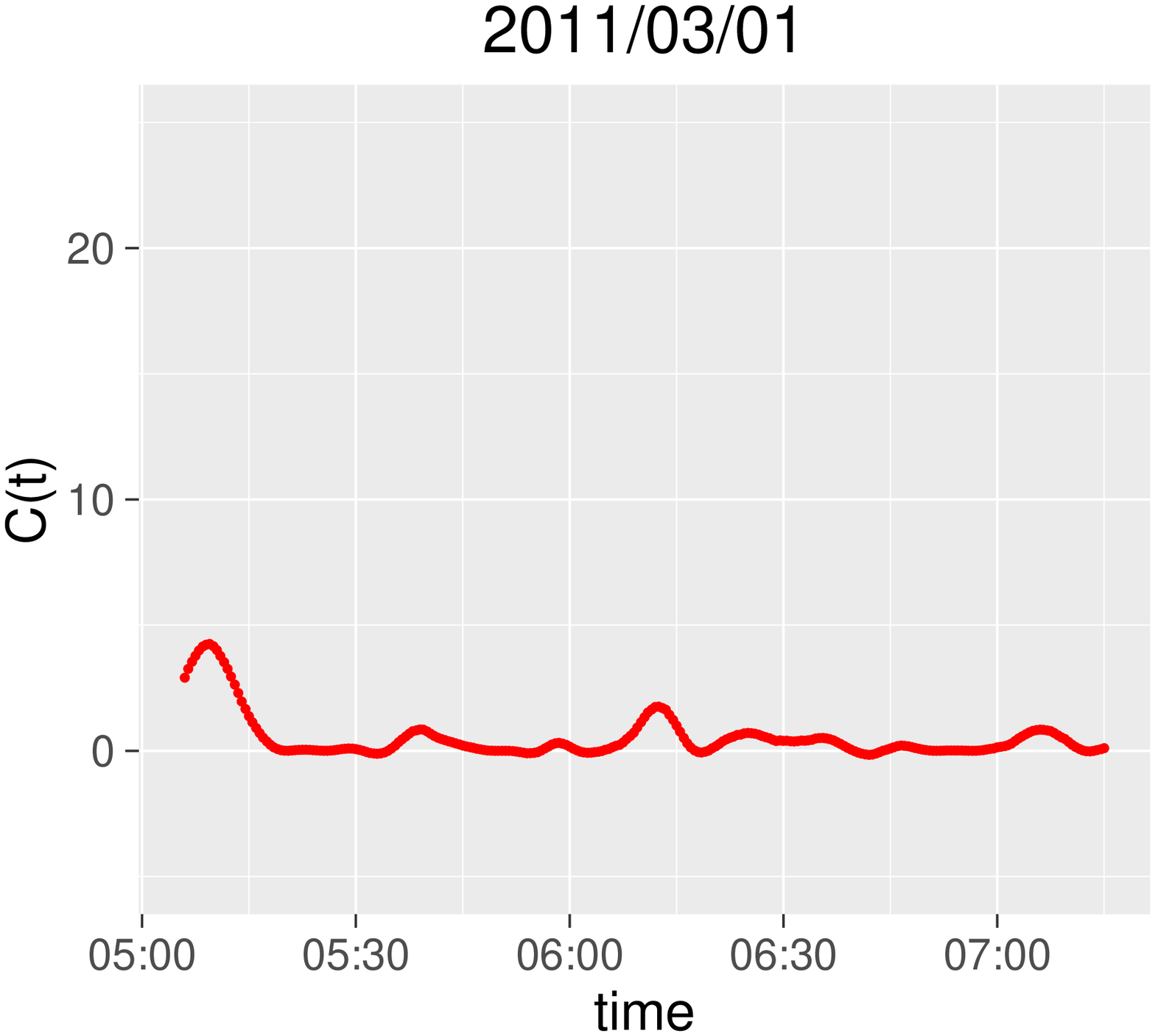}\\
\end{tabular}
\caption{The result of correlation analysis on non-earthquake days.  The vertical axis shows the correlation $C(T)$ and the horizontal one the time $t$ [UTC].  The day 40, 30, 20, 10 days before the earthquake, respectively.  We used 7th polynomial functions as fitting curves.}
\label{kitaibaraki30-60}
\end{figure}

\begin{figure}
\begin{tabular}{cc}
\noindent\includegraphics[width=14pc]{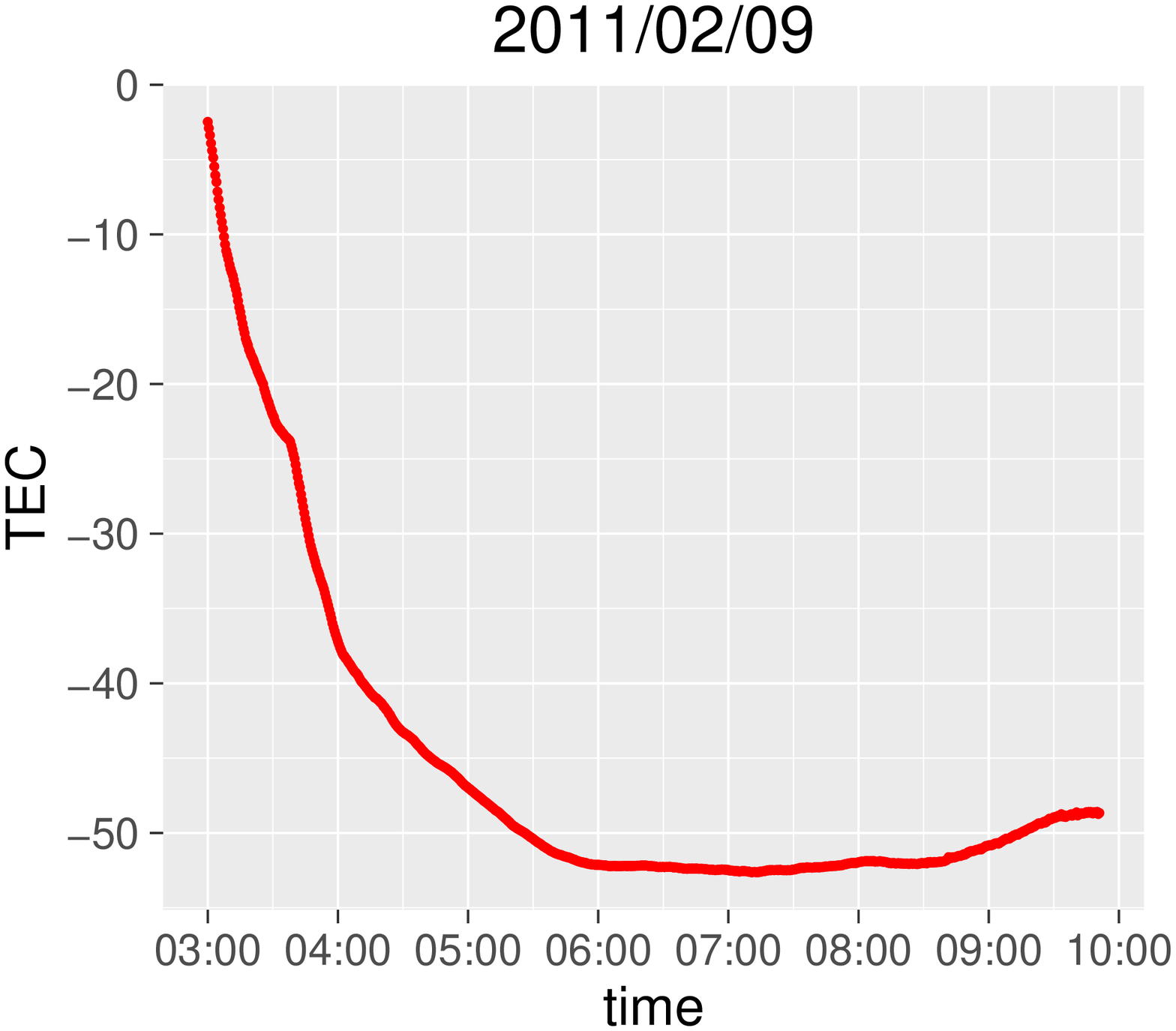} &
\noindent\includegraphics[width=14pc]{cor_0214_040_26poly7.eps}\\
\noindent\includegraphics[width=14pc]{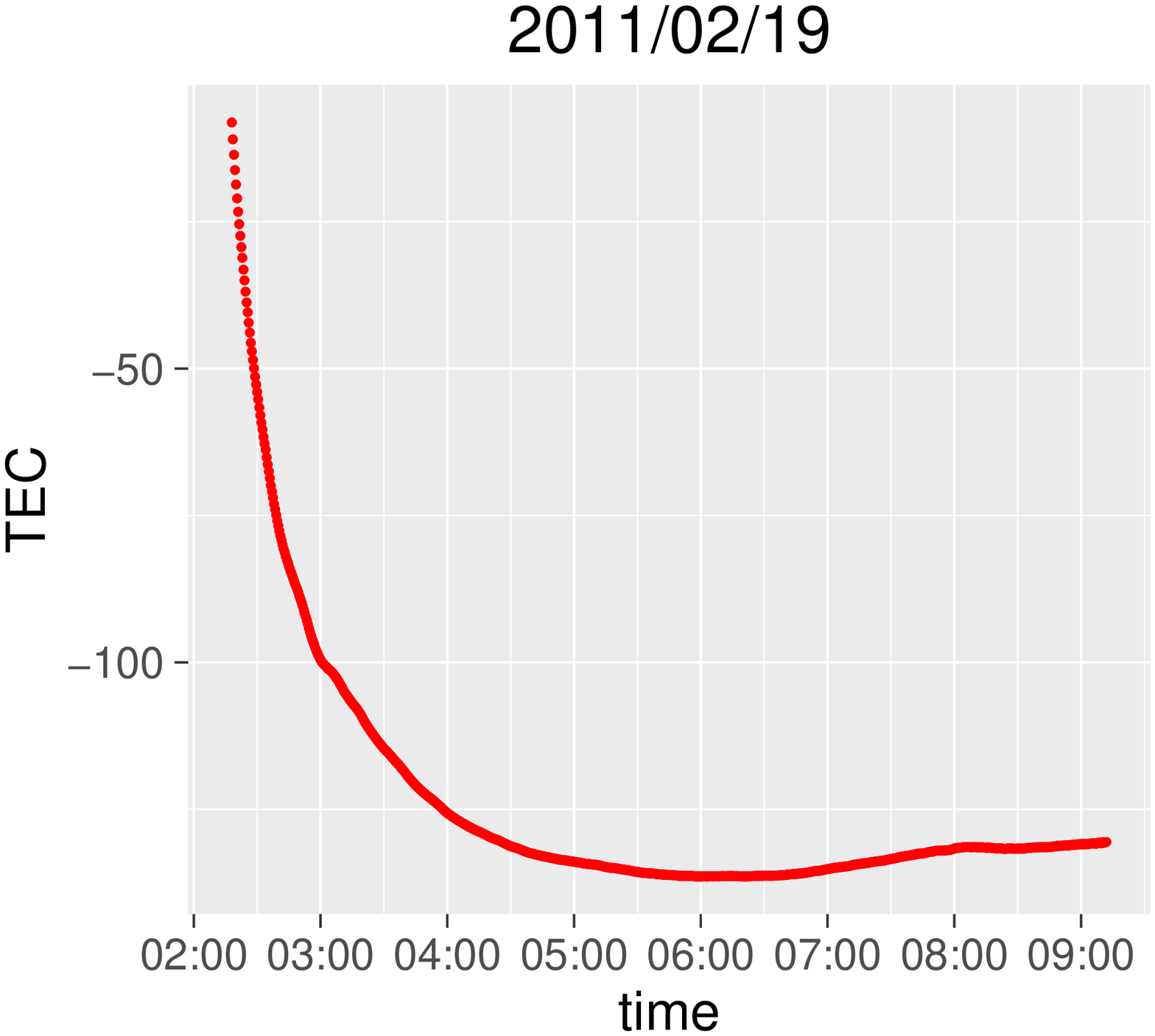} &
\noindent\includegraphics[width=14pc]{cor_0214_050_26poly7.eps}\\
\noindent\includegraphics[width=14pc]{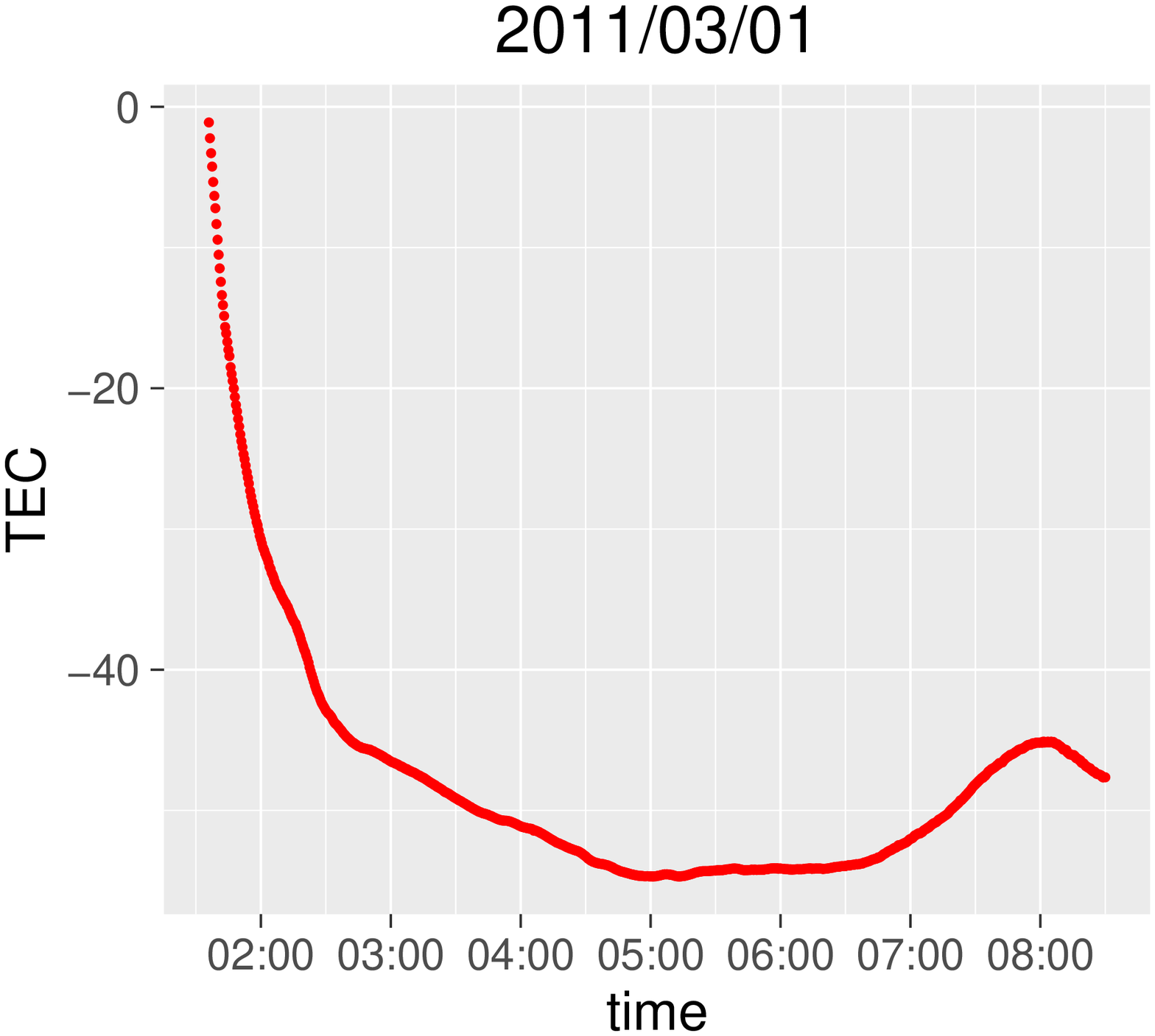} &
\noindent\includegraphics[width=14pc]{cor_0214_060_26poly7.eps}\\
\noindent\includegraphics[width=14pc]{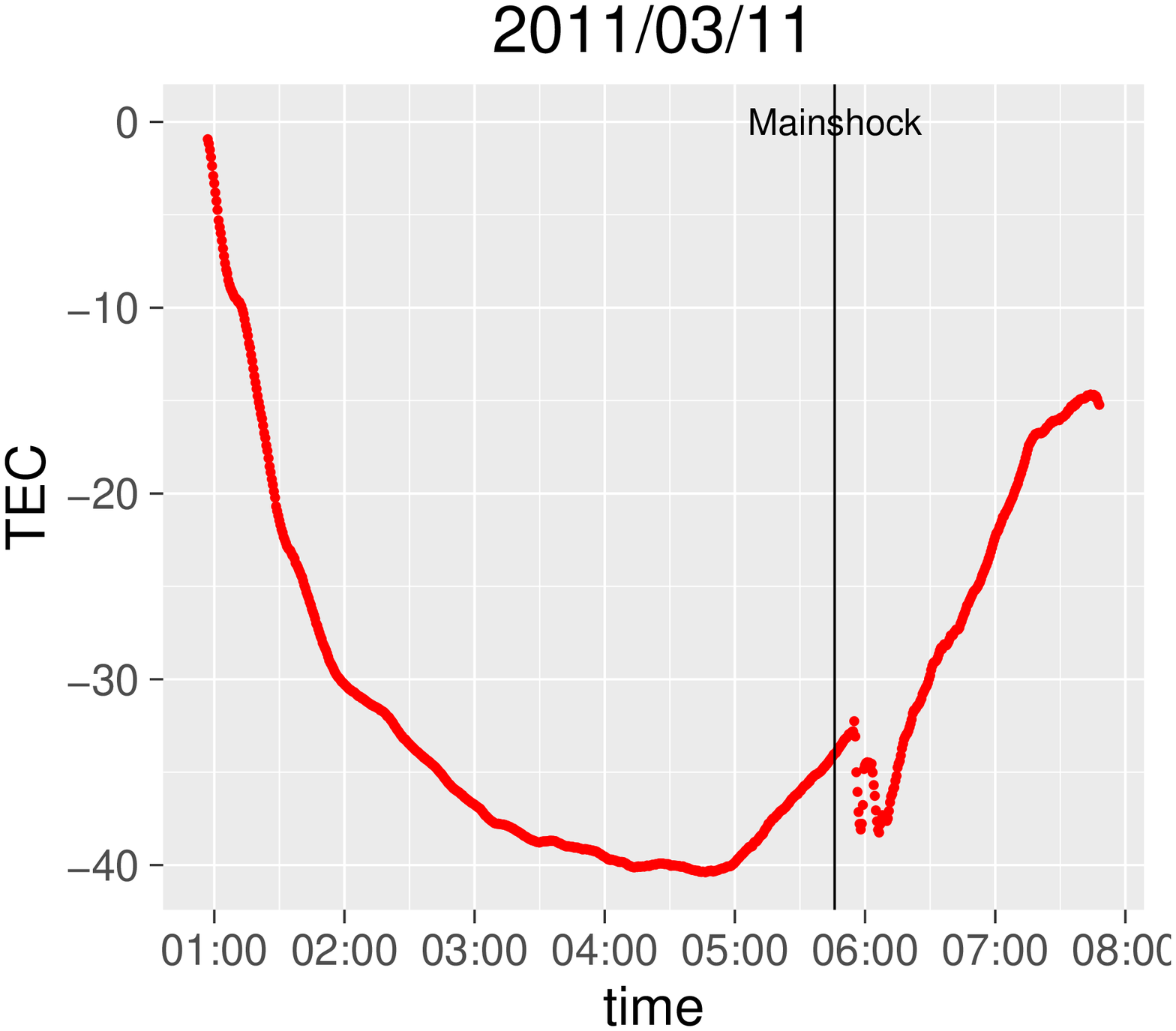} &
\noindent\includegraphics[width=14pc]{cor_0214_070_26poly7.eps}\\
\end{tabular}
\caption{The comparison between the real TEC (left) and the results of correlation analysis (right) on earthquake day and non-earthquake days.}
\label{tec-cor}
\end{figure}

\begin{figure}
\begin{tabular}{cc}
\noindent\includegraphics[width=18pc]{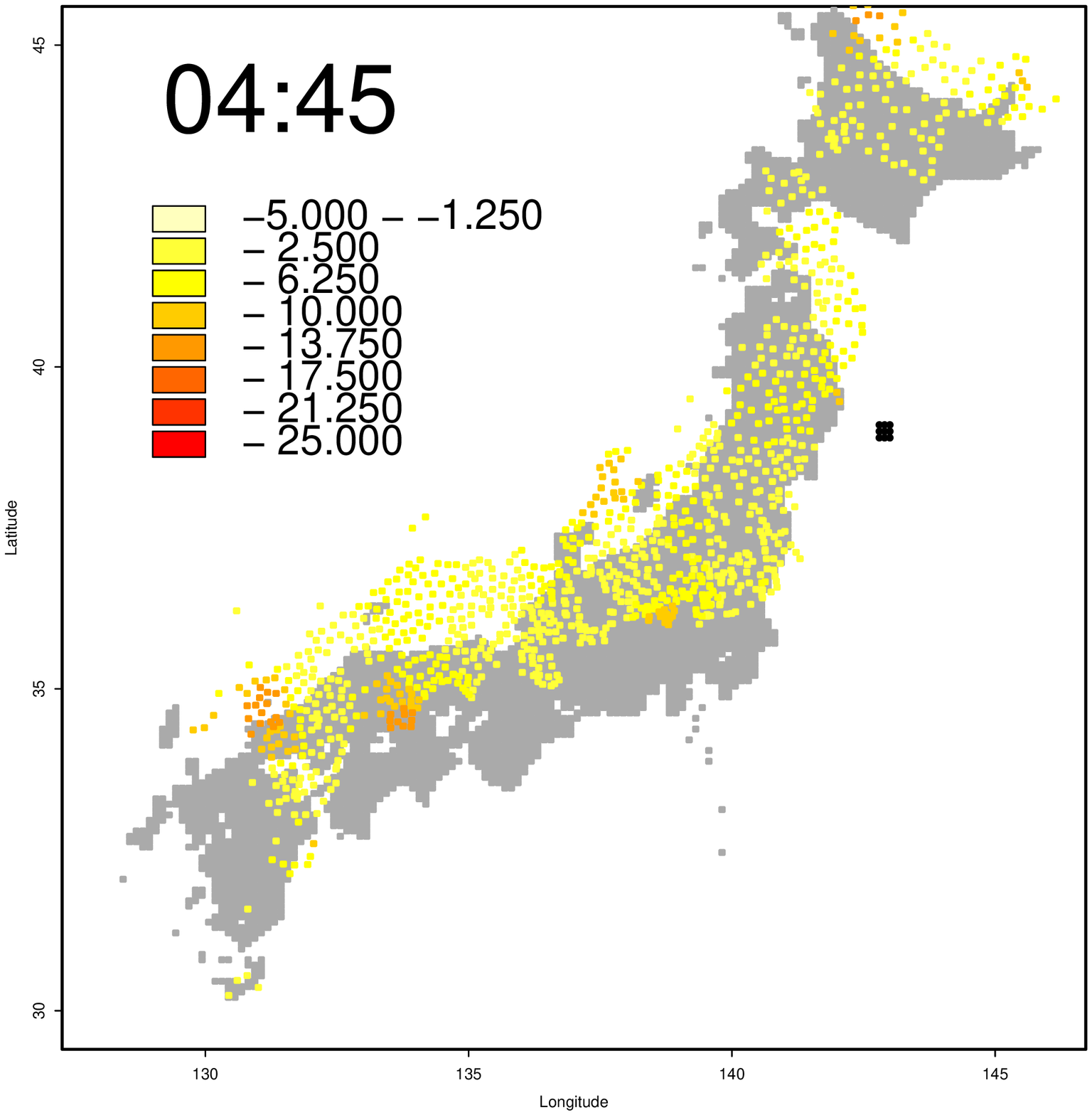} &
\noindent\includegraphics[width=18pc]{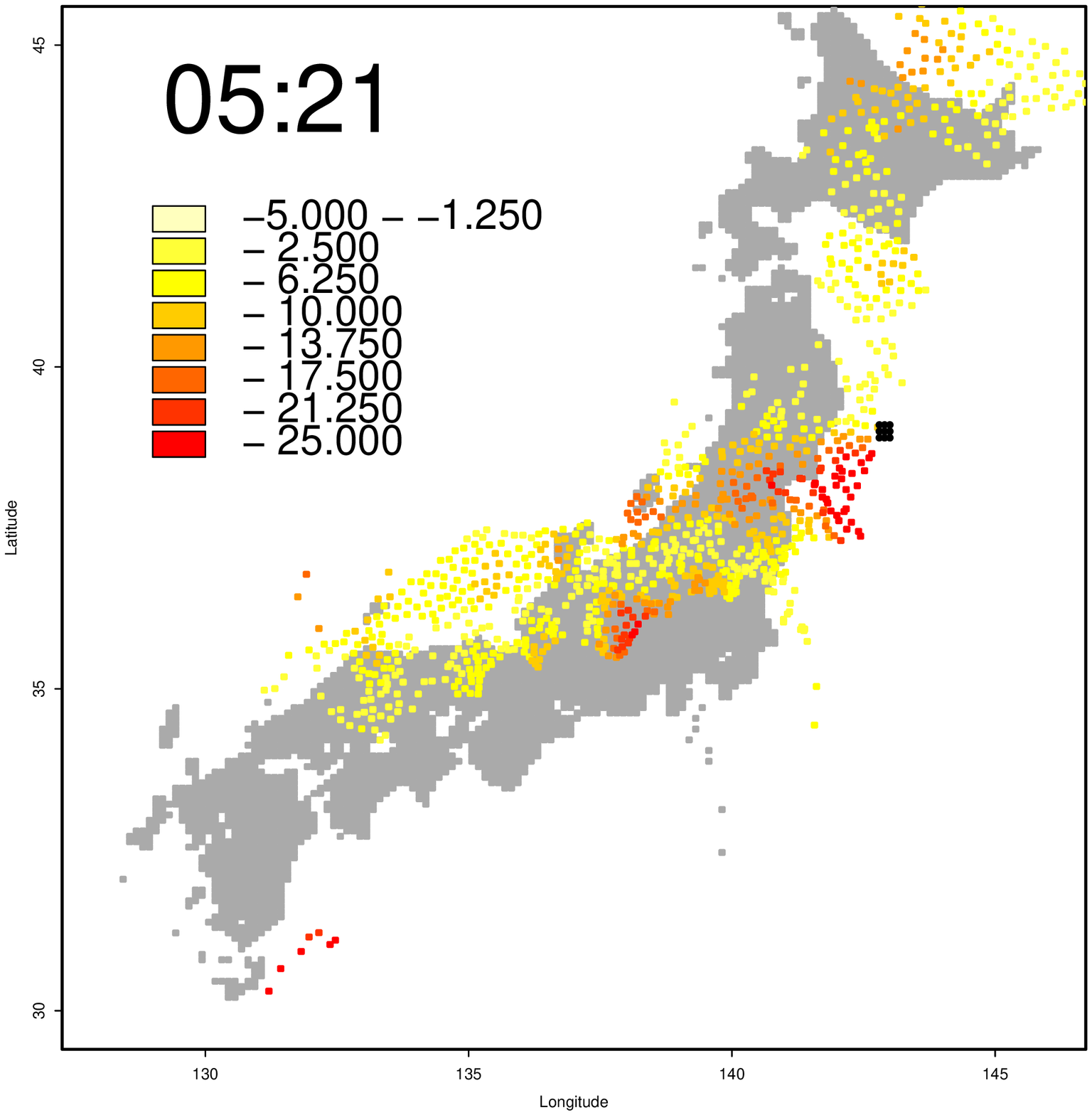}\\
 & \\
 & \\
\noindent\includegraphics[width=18pc]{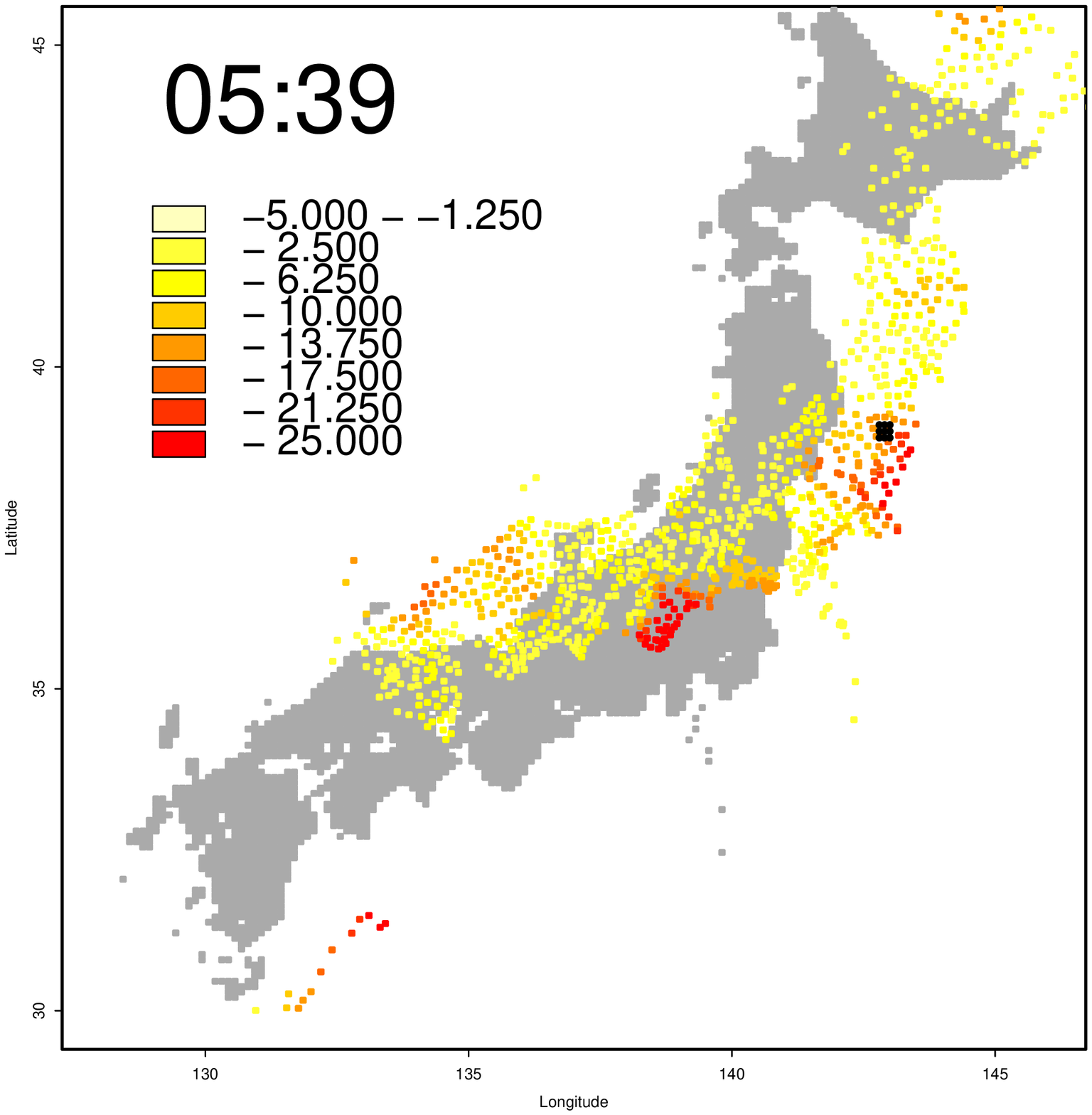} &
\noindent\includegraphics[width=18pc]{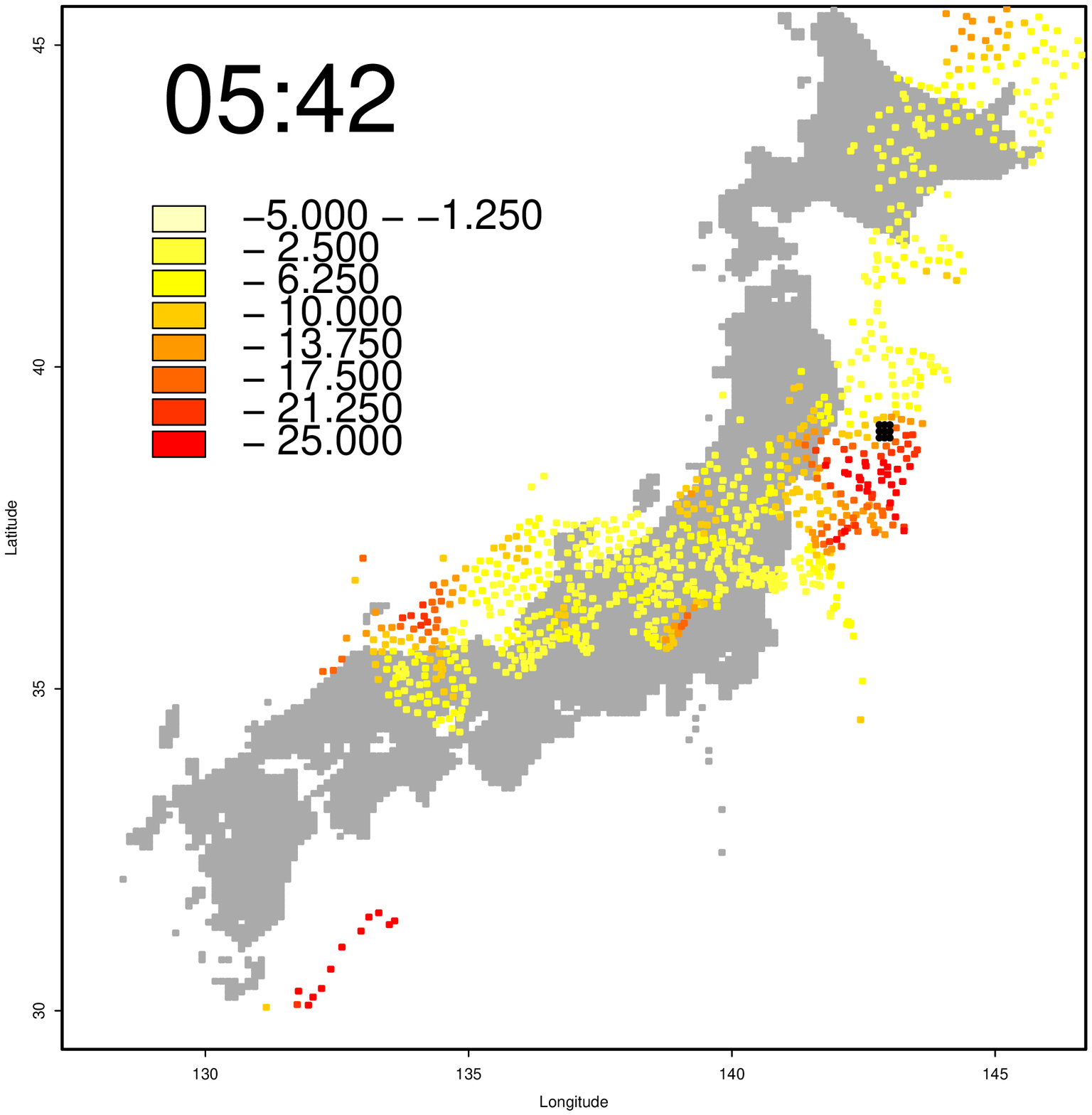} \\
\end{tabular}
\caption{The change of $C$($T$) at epochs shortly before the 2011 Tohoku-Oki earthquake, i.e. 1 hour, 25 minutes, 7 minutes, and 4 minutes before the actual time 05:46 [UTC] when the main shock of the 2011 Tohoku-Oki earthquate occured on March 11, 2011.  The black circle represents the epicenter.}
\label{japan70}
\end{figure}

\begin{figure}
\noindent\includegraphics[width=35pc]{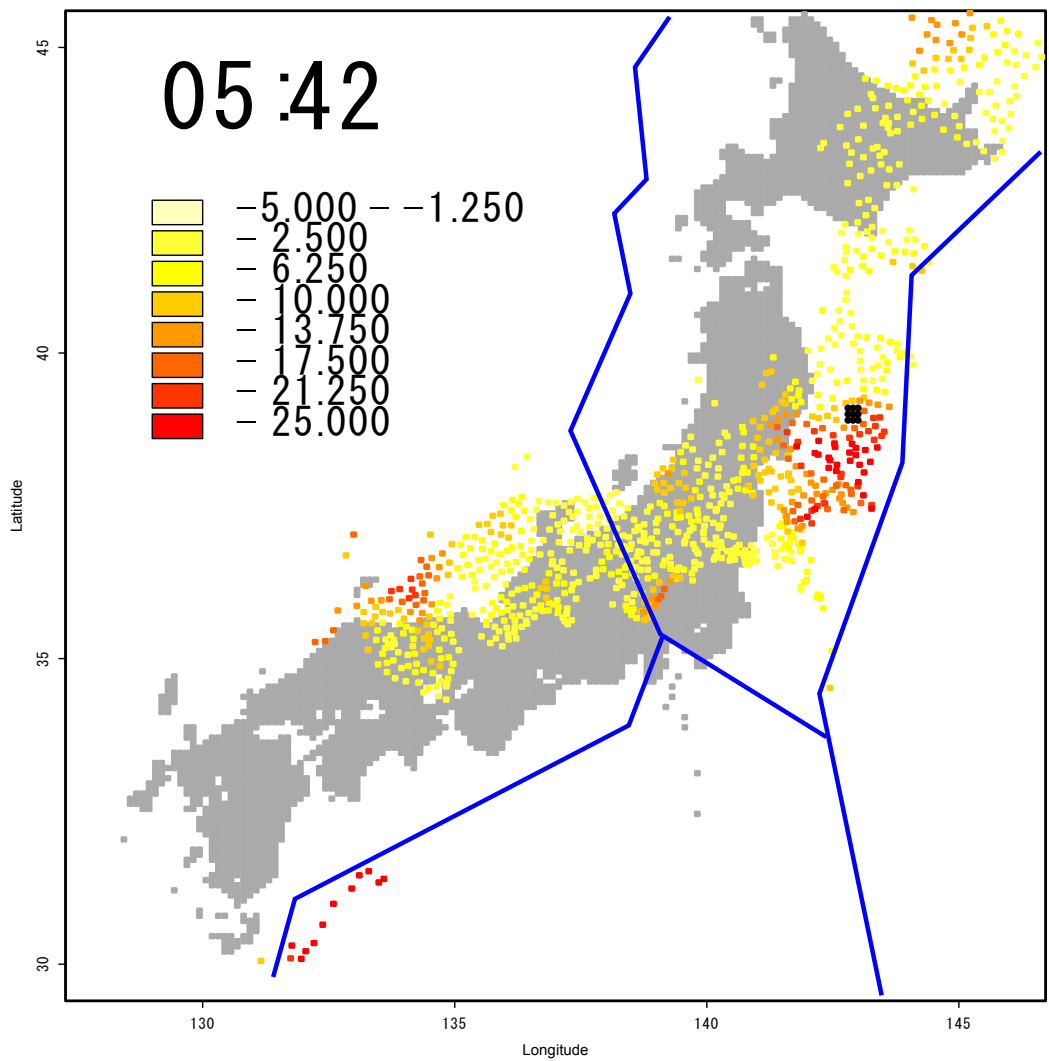}
\caption{The result of correlation analysis on 11th March, 2011.  The black circle represents the epicenter of the 2011 Tohoku-Oki earthquake.  The blue lines present the plate boundaries.}
\label{bound}
\end{figure}

\begin{figure}
\begin{tabular}{cc}
\noindent\includegraphics[height=10pc,width=15pc]{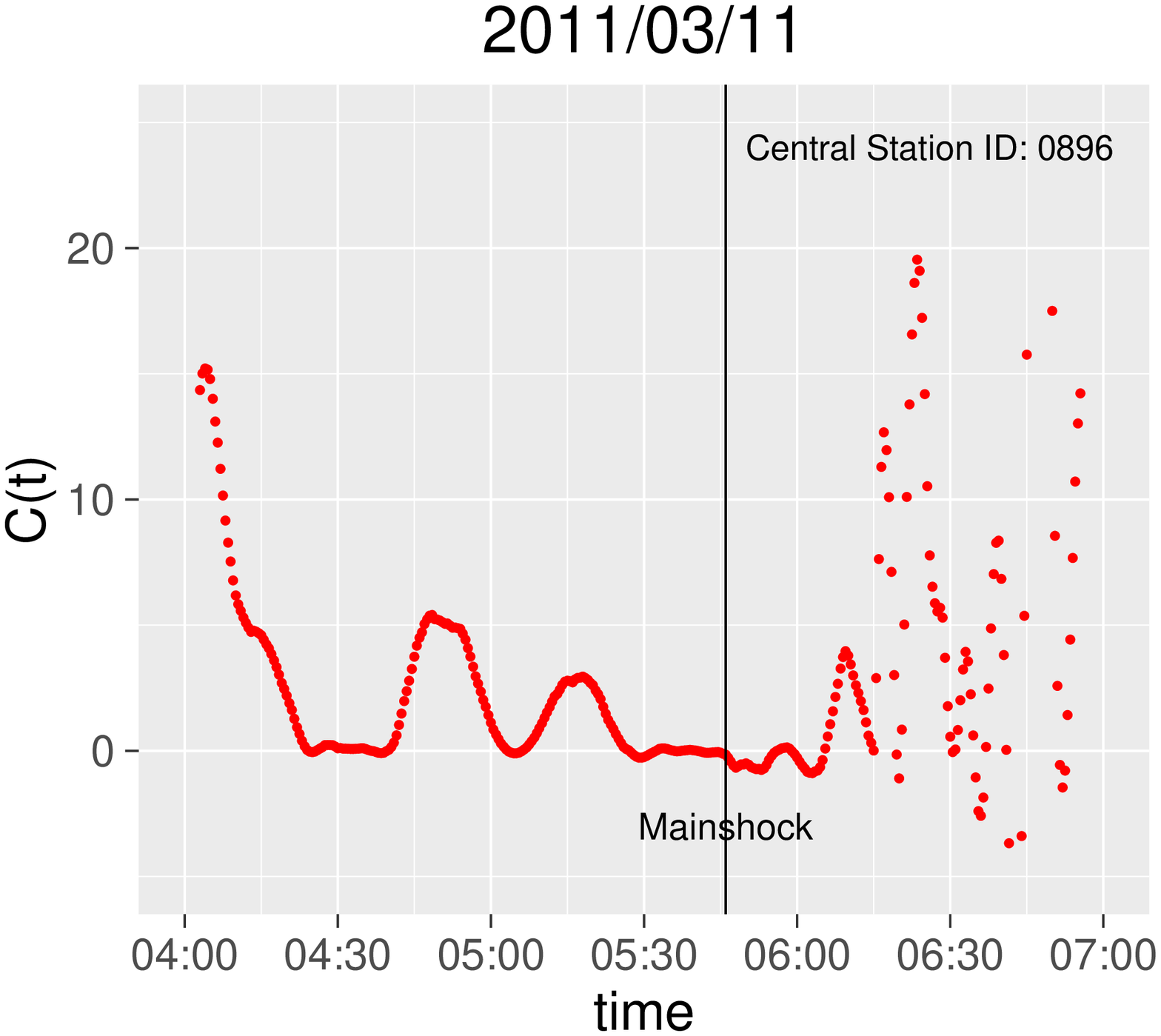} &
\noindent\includegraphics[height=10pc]{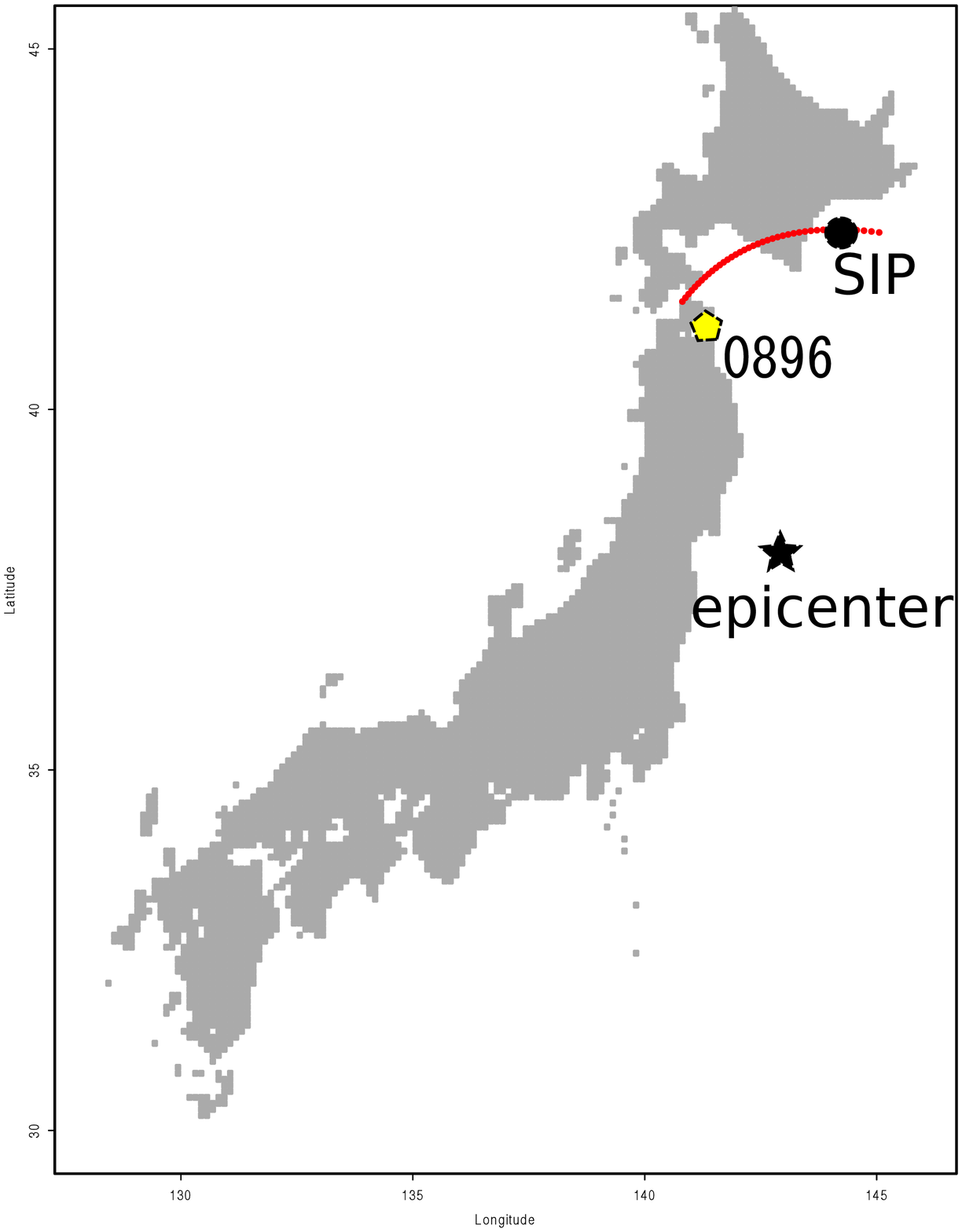}\\
\noindent\includegraphics[height=10pc,width=15pc]{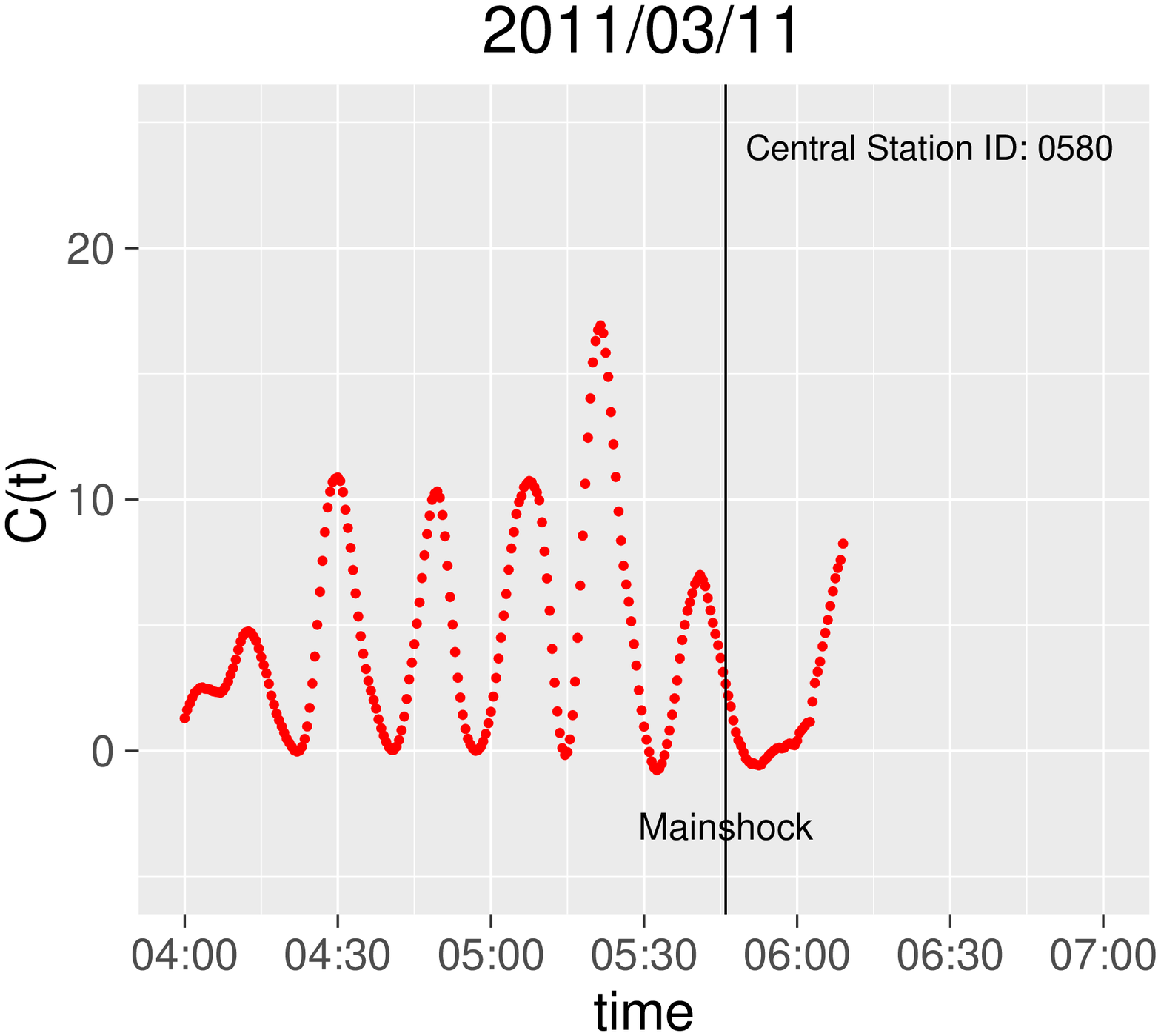} &
\noindent\includegraphics[height=10pc]{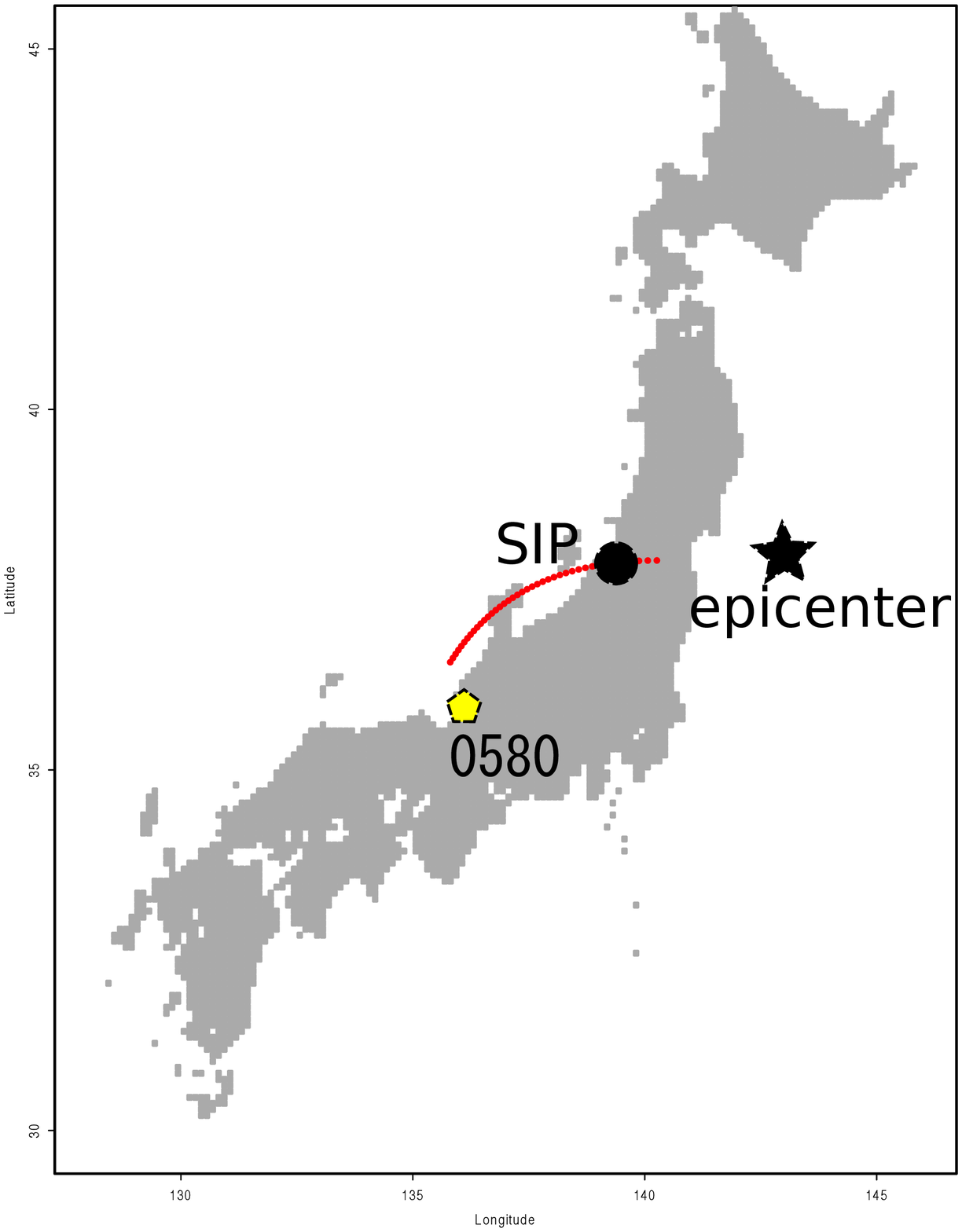}\\
\noindent\includegraphics[height=10pc,width=15pc]{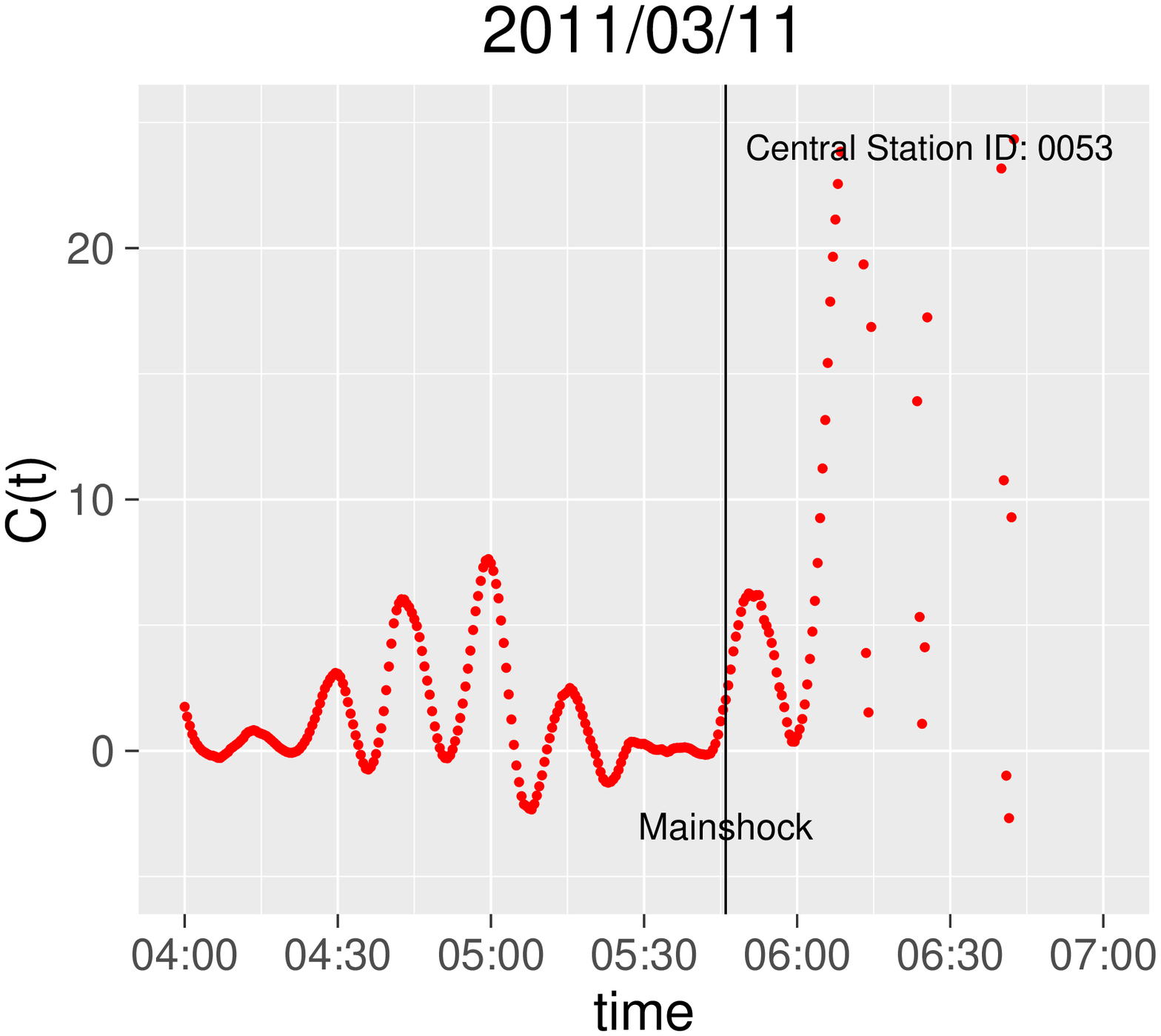} &
\noindent\includegraphics[height=10pc]{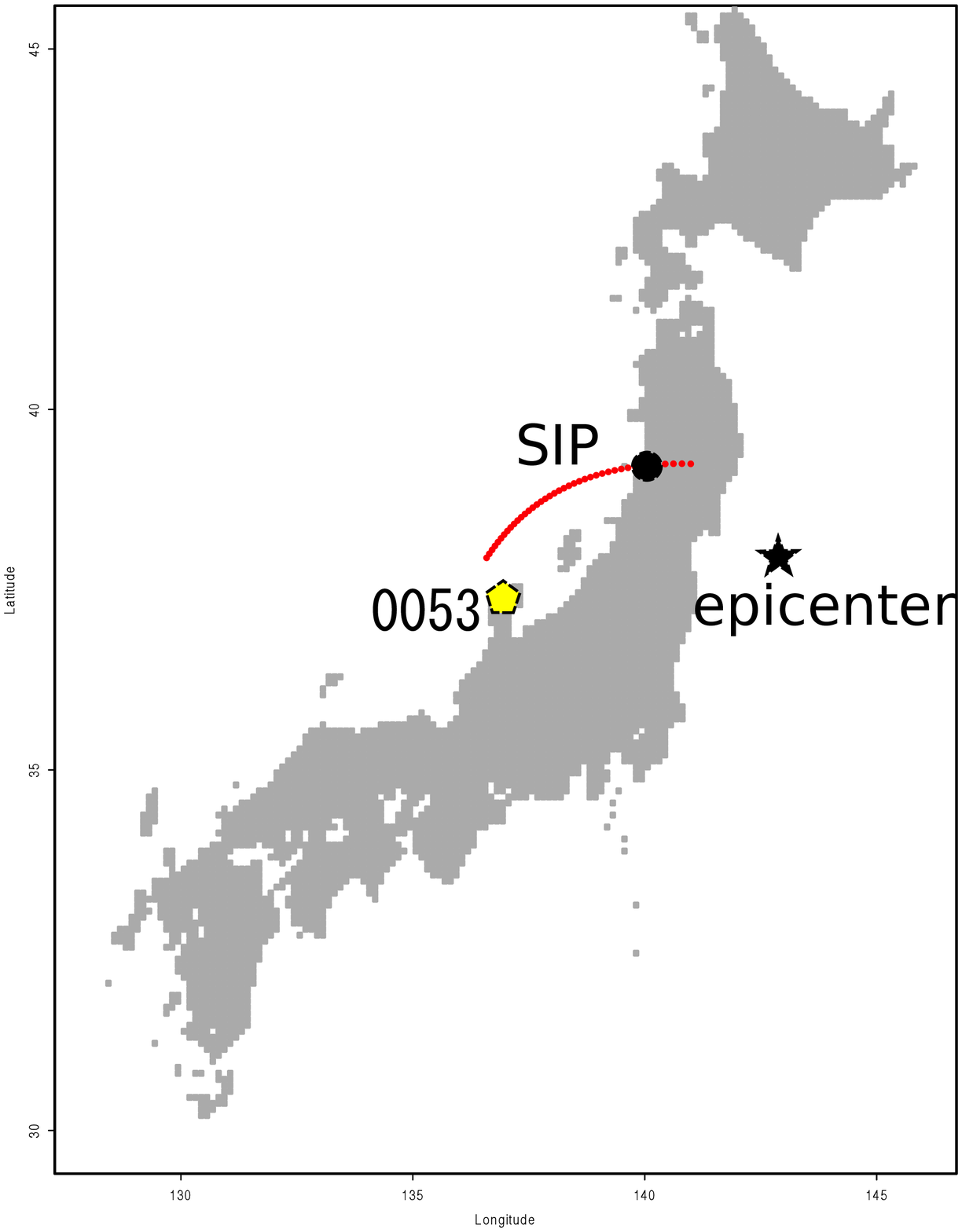}\\
\noindent\includegraphics[height=10pc,width=15pc]{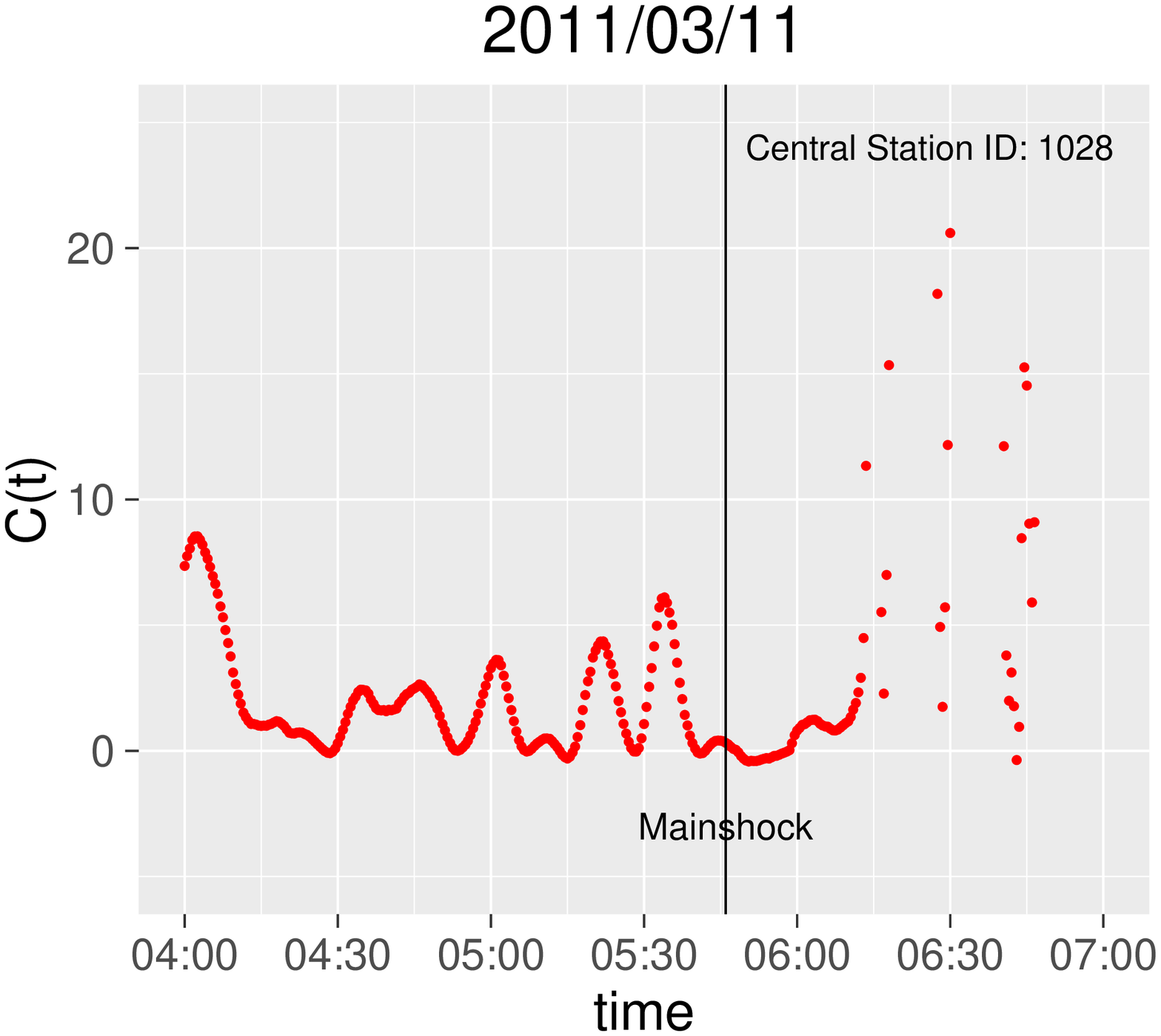} &
\noindent\includegraphics[height=10pc]{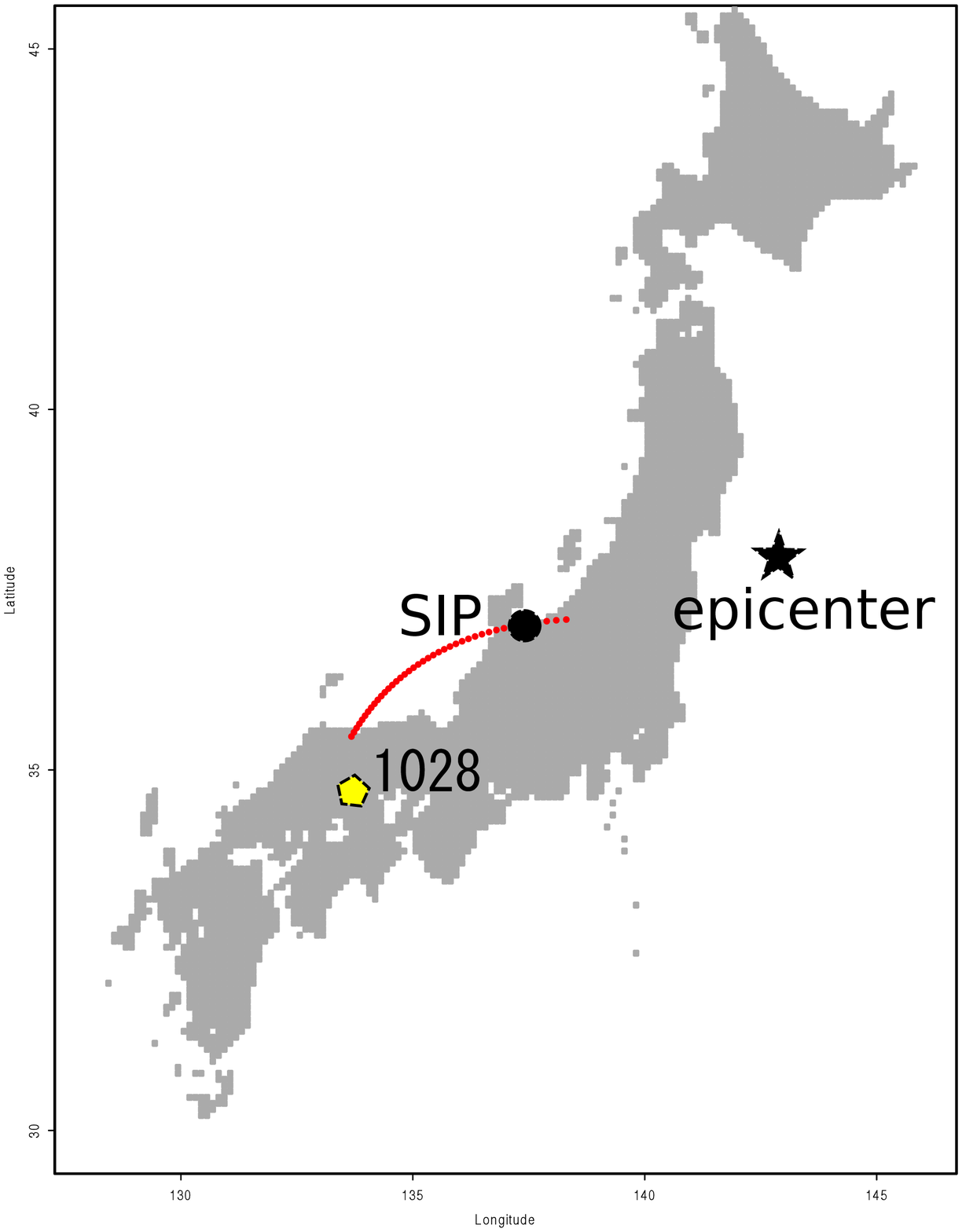}\\
\end{tabular}
\caption{The change of $C$($T$) on 11th March, 2011, at Aomori (0896), Fukui (0580), Ishikawa (0053), Okayama (1028) stations, and their positions in the map.  The vertical axis shows the correlation $C(T)$ and the horizontal one the time $t$ [UTC].  The red curves represent the tracks of the SIPs from 4:00 [UT] to 6:00 [UT], and the black points represent the SIPs when the earthquake occured.  The black lines indicate the exact time 05:46 [UTC] when the 2011 Tohoku-Oki earthquake occured.}
\label{other70}
\end{figure}

\begin{figure}
\noindent\includegraphics[width=35pc]{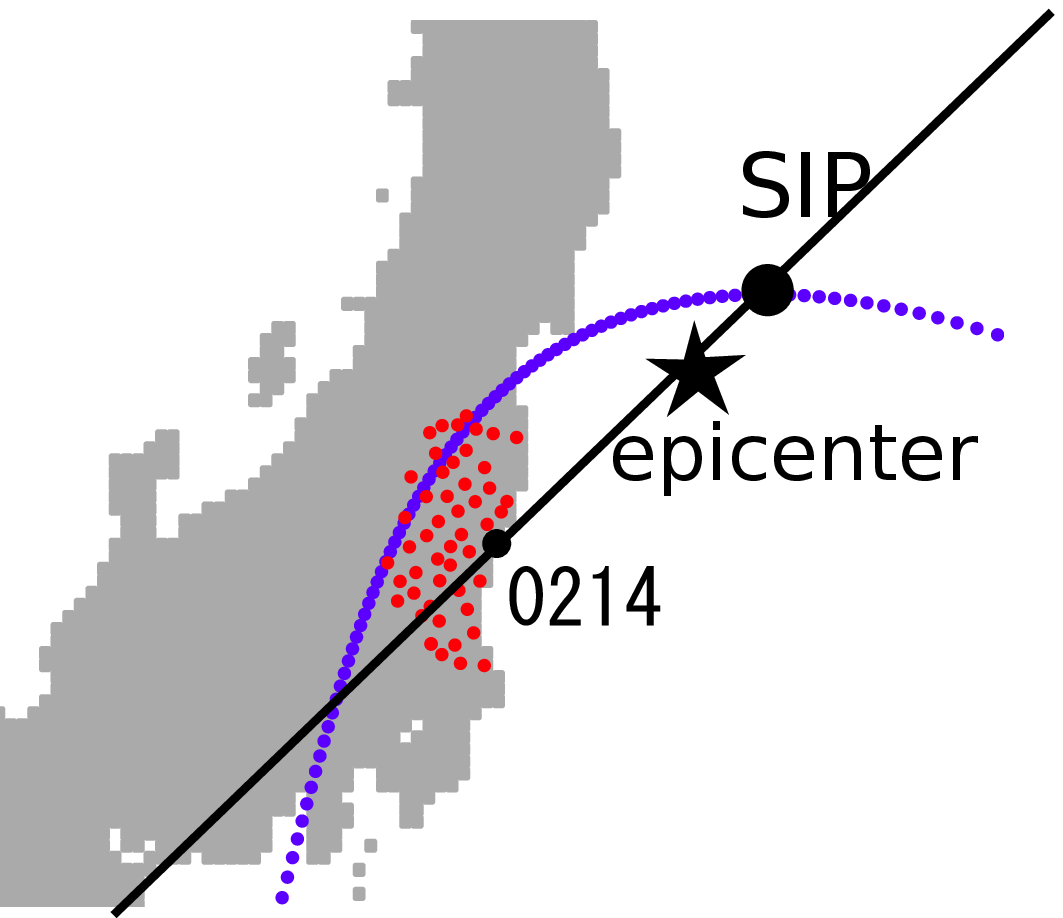}
\caption{The blue line represents the SIP track of the pair of the 0214 station (Kitaibaraki) and the GPS satellite 26, and the red points represent the location of surrounding 50 stations around the central station.  The star represents the epicenter of the 2011 Tohoku-Oki earthquake, and the black circle on the blue line represents the SIP position at the earthquake occurrence time.}
\label{siptrack}
\end{figure}

Figure \ref{other70} shows the change of $C$($T$) and their locations at four other stations on the earthquake day.
In comparison to Fig. \ref{kitaibaraki70}, $C$($T$) are relatively small at the four stations.
Note that here $C$($T$) at Fukui seems to be partially anomalous, which is similar to Fig. \ref{kitaibaraki70}.
We think that such partial anomaly of $C$($T$) at Fukui is observed because our correlation analysis can detect "anomaly" and its SIP of 0580 Fukui station was \textit{near} the epicenter of the 2011 Tohoku-Oki earthquake.\\
Hence, we used the GPS satellite 26.
There were other satellites above Japan at this time, however, adequate enough length of TEC data before the earthquake is required to observe preseismic ionospheric condition.
We think that the GPS satellite 26 had such appropriate length of TEC data because the track line between the SIP and the station is near the epicenter of the 2011 Tohoku-Oki earthquake as shown in Fig.\ref{siptrack}.

\subsection{Dependance on fitting functions}
Figure \ref{compare70} shows the comparison of results of correlation analysis with different extrapolating functions used in STEP 2.\\
We used 7-th polynomial functions, 5-th polynomial functions, 3rd Fourier series and 7-th Gaussian functions.
It is seen that the results largely depend on the fitting functions.
However, the anomalous trend before the earthquake can be seen in every case.\\
Figures \ref{kitaibaraki30-60p5},\ref{kitaibaraki30-60f3} and \ref{kitaibaraki30-60g7} show the results of correlation analysis on non-earthquake days with different fitting functions.
The scale of the vertical axis vary with the fitting functions.
However, we can confirm that the correlation values on the earthquake day are anomalous as compared with non-earthquake days whichever the fitting functions we use.

\subsection{Physical Mechanism}
The physical mechanism of the anomaly has been researched so far and several models which explain the preseismic ionospheric anomalies are introduced \cite{Kuo2011,Kuo2014}.
As for the 2011 Tohoku-Oki case, the preseismic anomaly is simulated by using these models \cite{Kuo2015}, where, the electric coupling model is used and the simulation results suggest that this model can explain the mechanism of the ionospheric anomalies before large earthquakes.
To connect such anomaly with the physical mechanism, we think that more studies to define an "appropriate anomaly" based on careful analysis.

\section{Conclusion}
In this paper, we introduced correlation analysis method into TEC data analysis.
We detected the TEC anomalies about \textit{one hour before} the 2011 Tohoku-Oki earthquake by applying this analysis method.
The TEC anomalies just about \textit{one hour before} the earthquake showed characteristic patterns which do not depend on our choice of extrapolation functions for nonlinear regression.
Although we still do not know the actual physical mechanism responsible for it, the present study based on the correlation analysis could be used for a link between the anomalies observed and the physical mechanism.
Further investigation of other large earthquakes such as earthquakes in Chile and understanding of physical mechanism of the anomaly should be pursued for practical application of correlation analysis to detect preseismic anomalies of potential great earthquakes such as the 2011 Tohoku-Oki earthquake on 11th March, 2011.

\section{Acknowledgements}
The GPS data have been downloaded from the Geospatial Information Authority of Japan (www.terras.gsi.go.jp).

\begin{figure}
\begin{tabular}{cc}
\noindent\includegraphics[height=10pc,width=15pc]{cor_0896_070_26poly7.eps} &
\noindent\includegraphics[height=10pc]{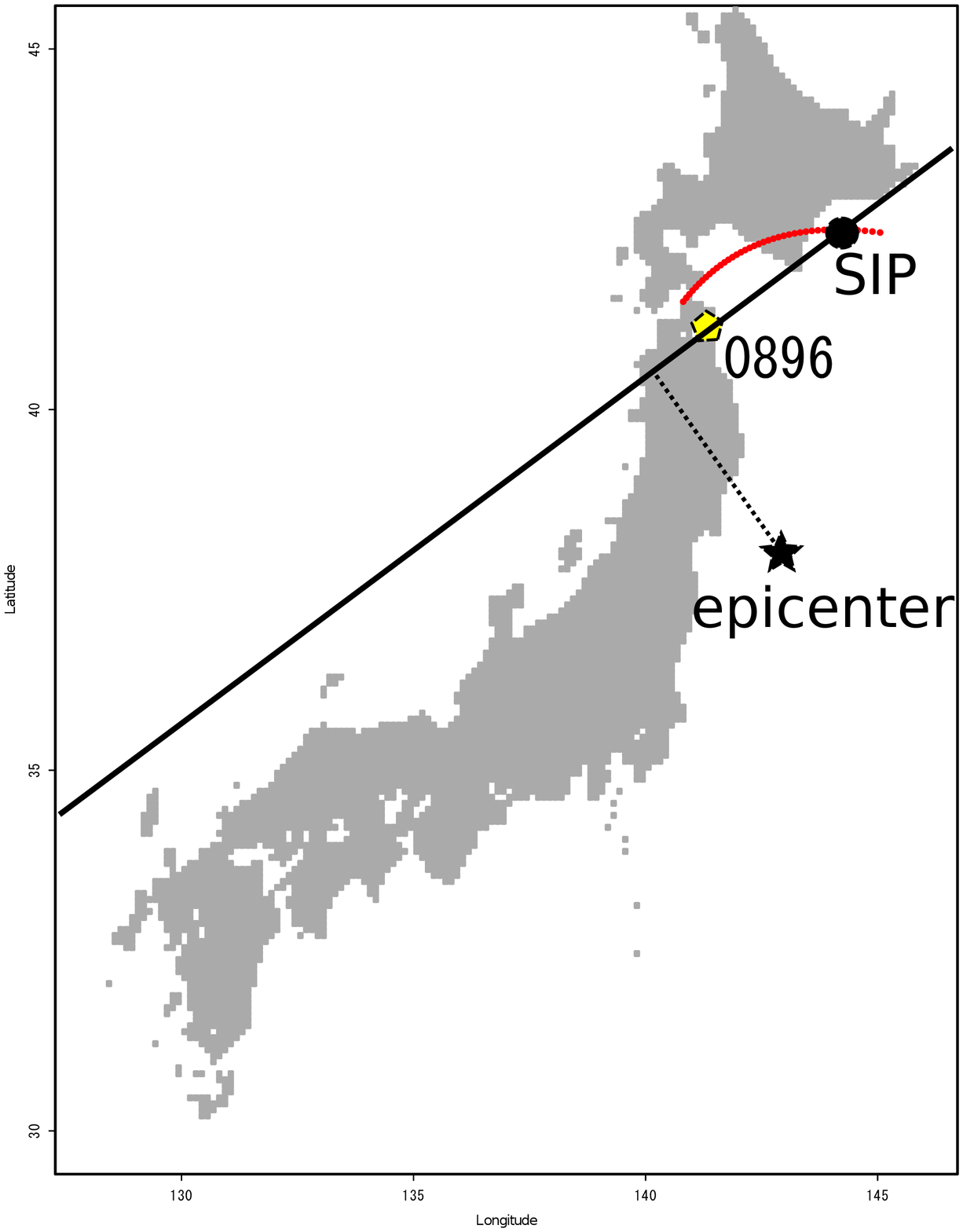}\\
\noindent\includegraphics[height=10pc,width=15pc]{cor_0580_070_26poly7.eps} &
\noindent\includegraphics[height=10pc]{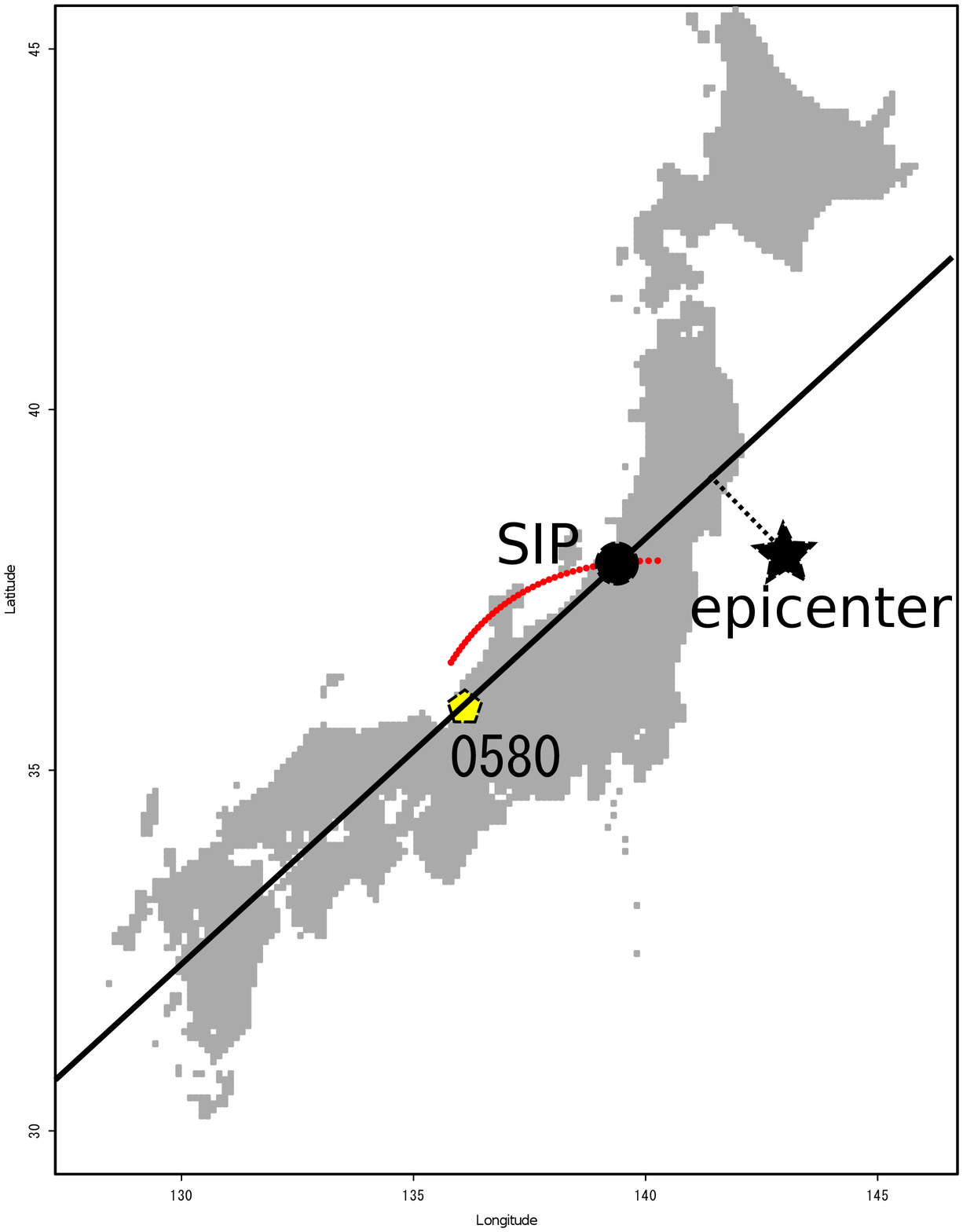}\\
\noindent\includegraphics[height=10pc,width=15pc]{cor_0053_070_26poly7.eps} &
\noindent\includegraphics[height=10pc]{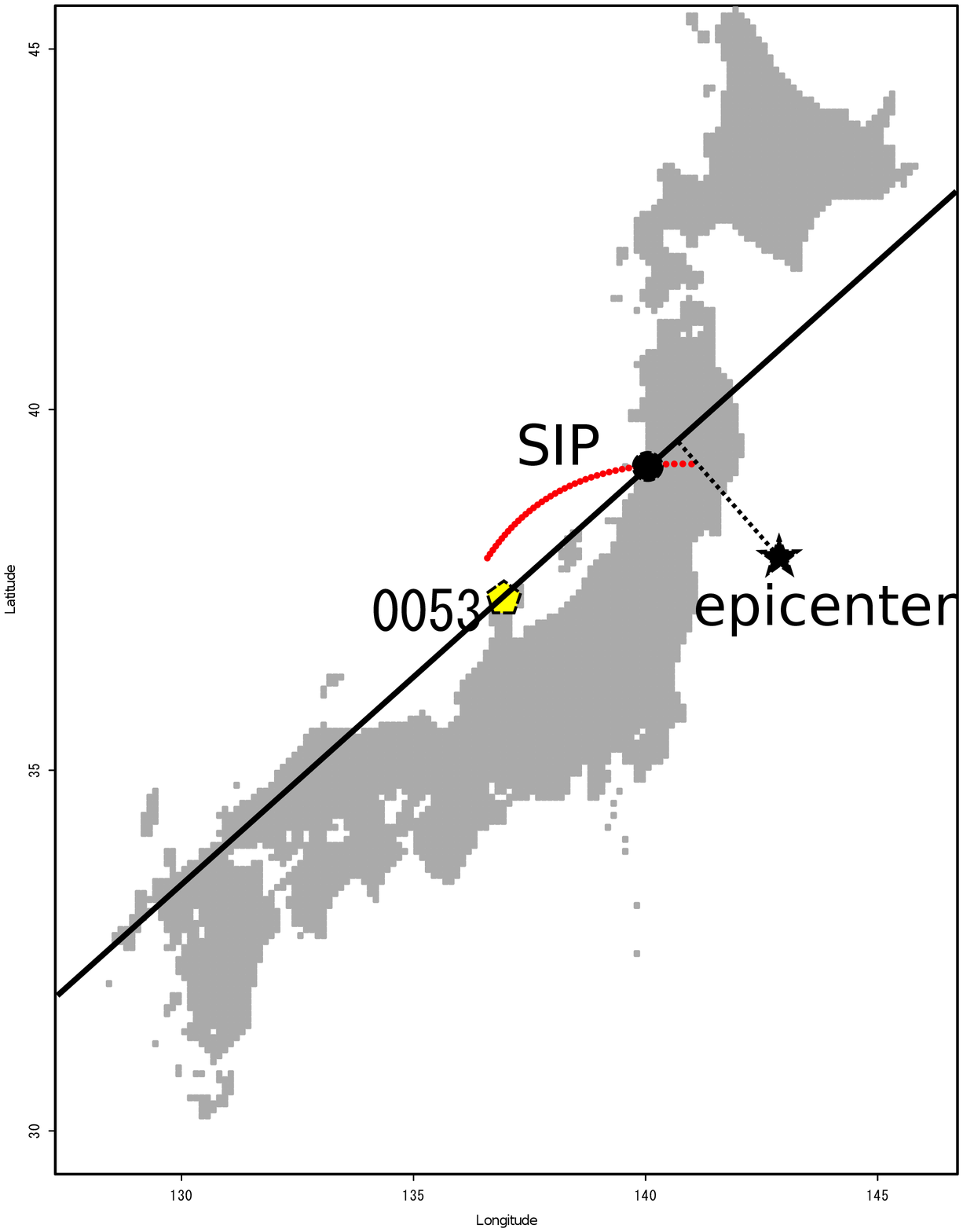}\\
\noindent\includegraphics[height=10pc,width=15pc]{cor_1028_070_26poly7.eps} &
\noindent\includegraphics[height=10pc]{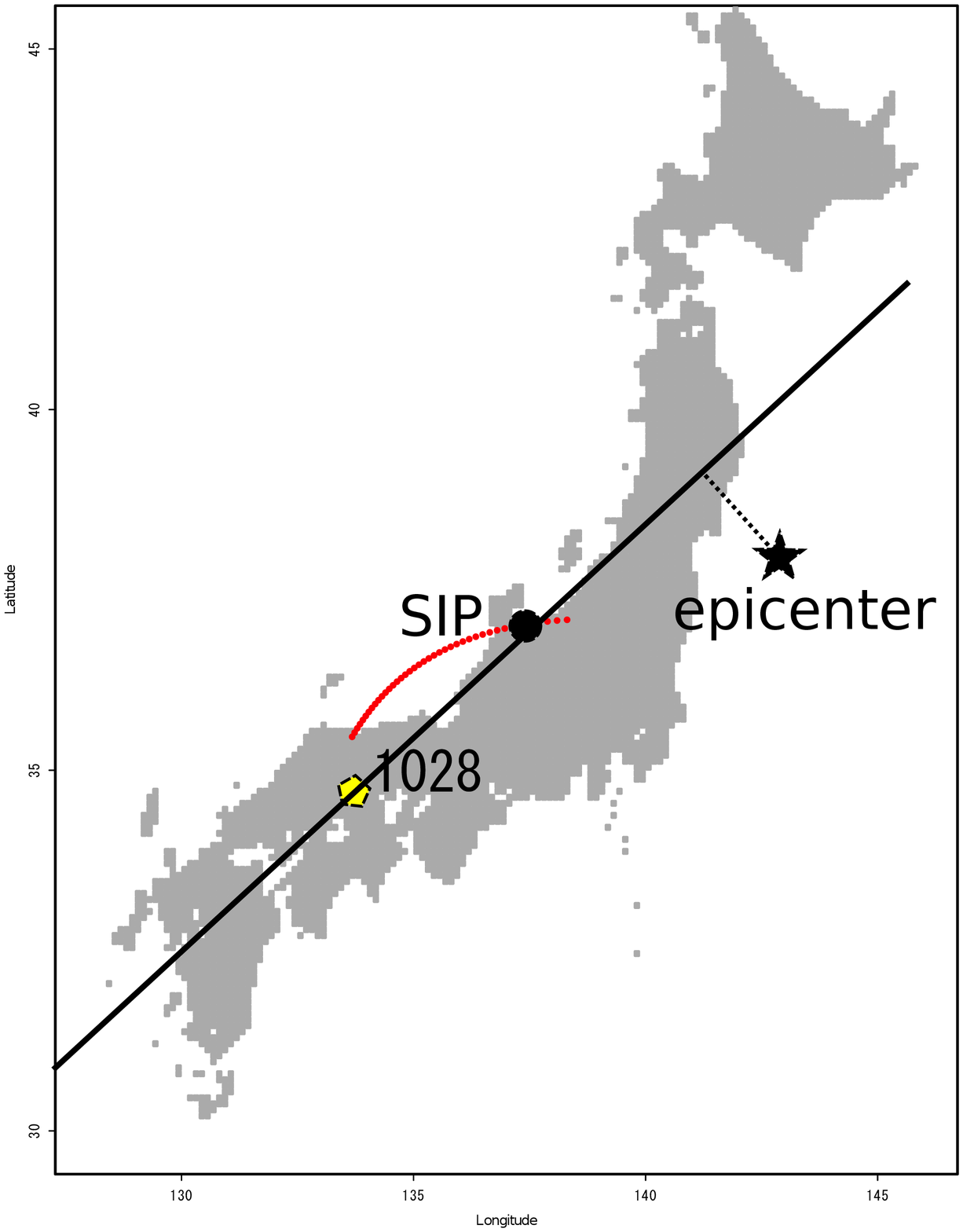}\\
\end{tabular}
\caption{The change of $C$($T$) on 11th March, 2011, at Aomori (0896), Fukui (0580), Ishikawa (0053), Okayama (1028) stations, and their positions in the map.  The vertical axis shows the correlation $C(T)$ and the horizontal one the time $t$ [UTC].  The red curves represent the tracks of the SIPs from 4:00 [UT] to 6:00 [UT], and the black points represent the SIPs when the earthquake occured.  The black lines indicate the exact time 05:46 [UTC] when the 2011 Tohoku-Oki earthquake occured.}
\label{other70track}
\end{figure}
\begin{figure}
\begin{tabular}{cccc}
(a) & &(b)& \\
\noindent(c) & \includegraphics[height=15pc]{cor_0214_070_26poly7.eps} &
\noindent(d) & \includegraphics[height=15pc]{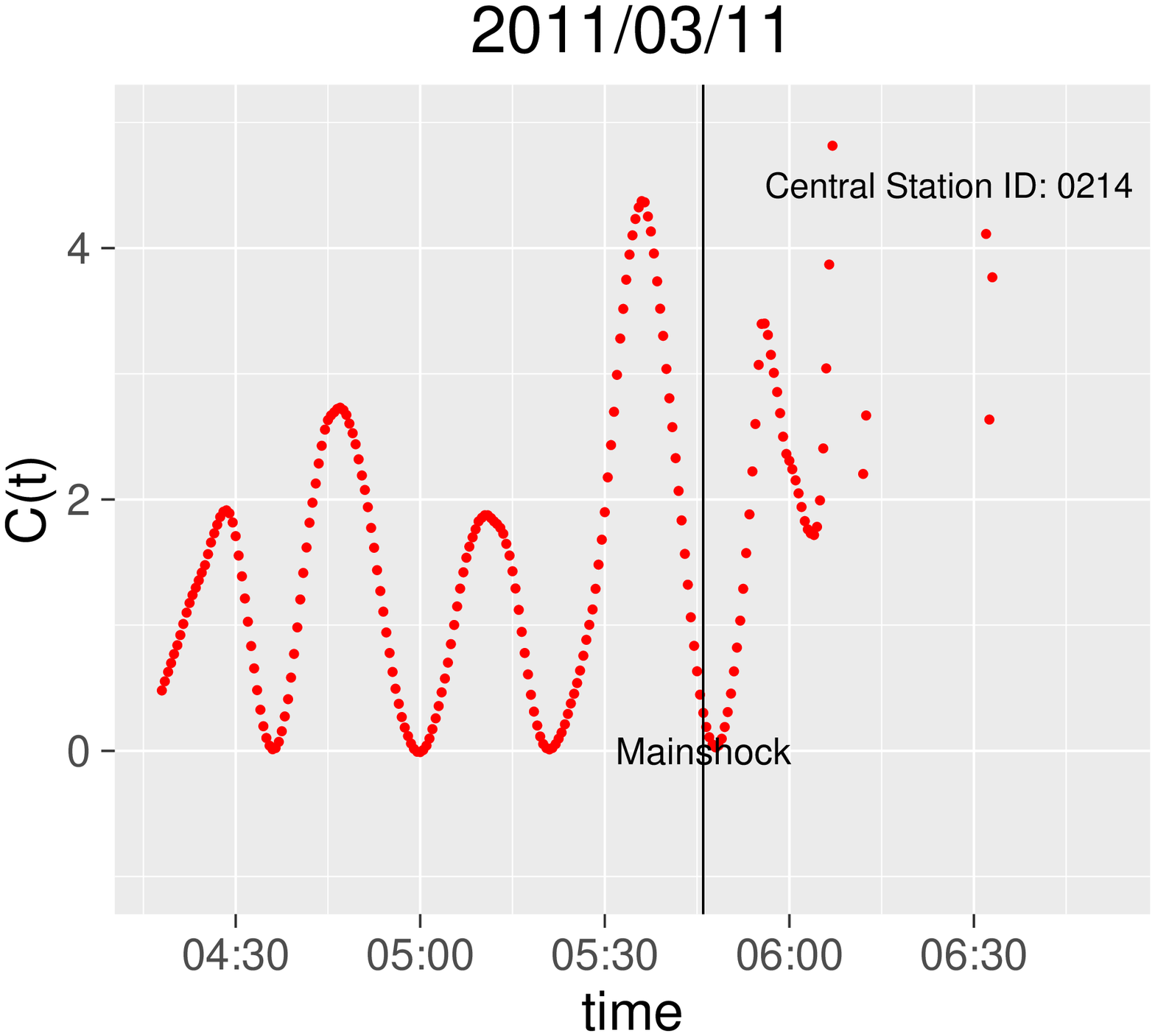}\\
\noindent& \includegraphics[height=15pc]{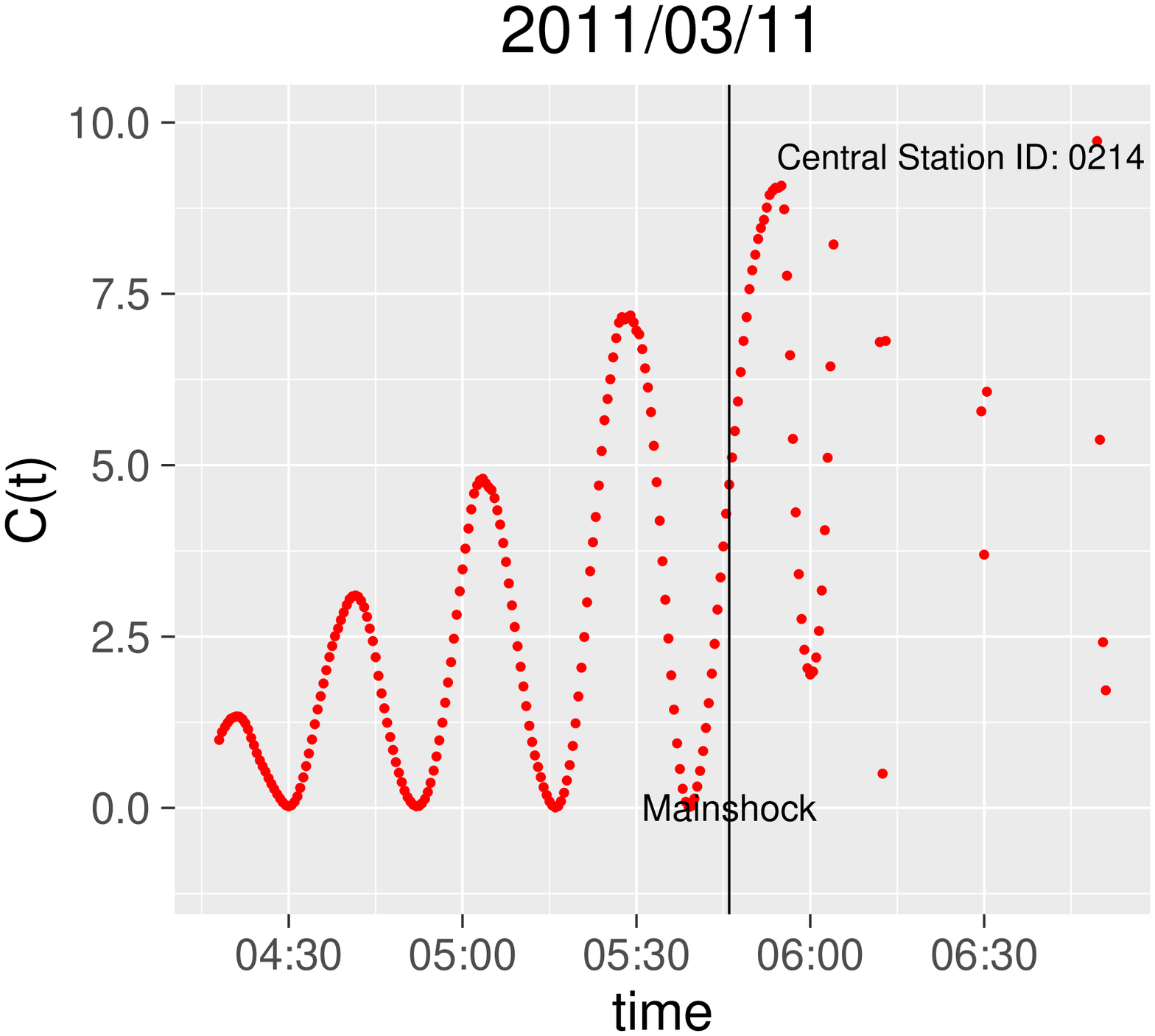} &
\noindent& \includegraphics[height=15pc]{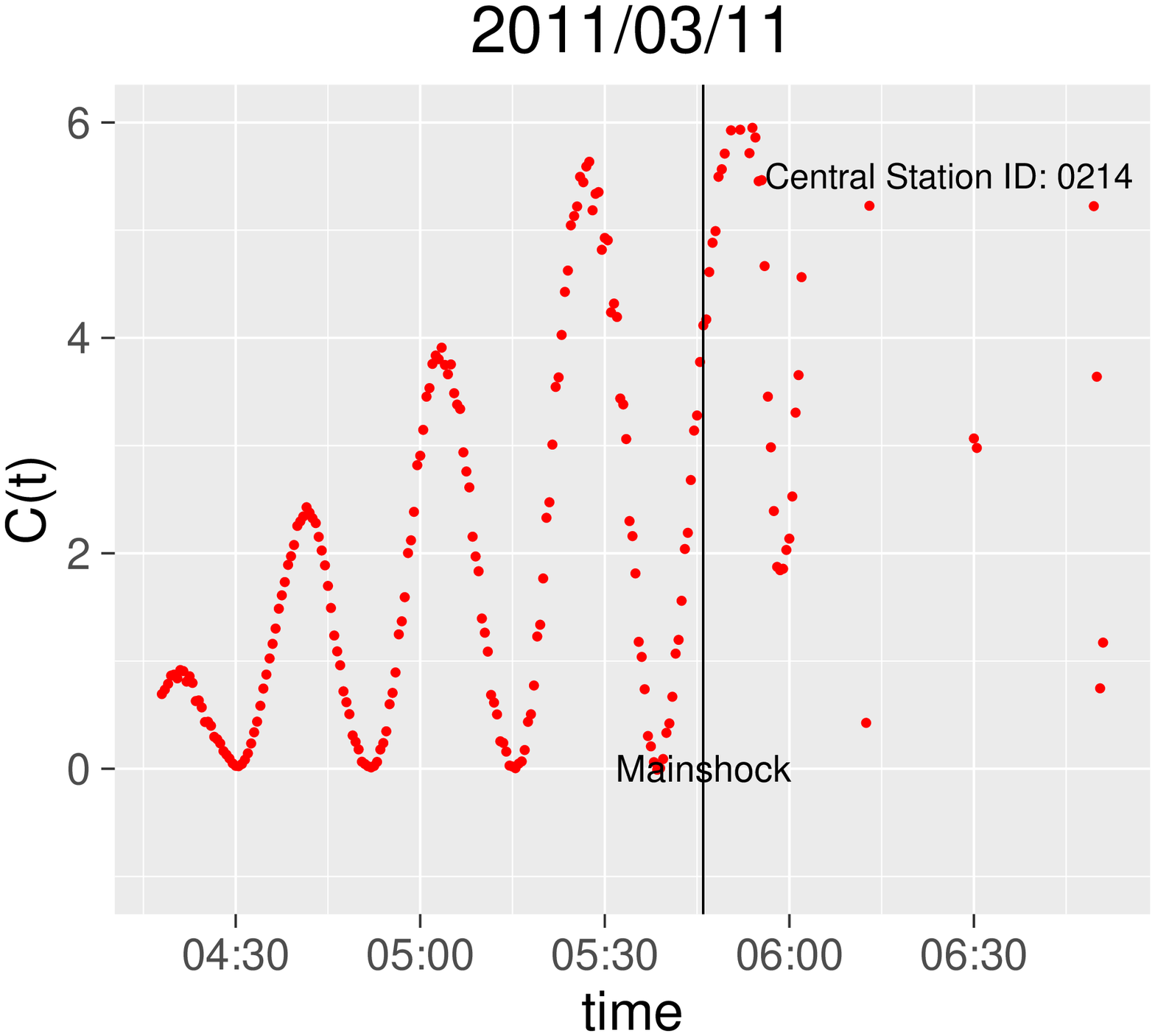}\\
\end{tabular}
\caption{The comparison of results of correlation analysis with different functions on the earthquake day.  0214 station and satellite 26 are used.  (a) 7-th polynomial function (b) 5-th polynomial function (c) 3rd Fourier series (d) 7-th Gaussian function.  The black lines indicate the exact time 05:46 [UTC] when the 2011 Tohoku-Oki earthquake occured.}
\label{compare70}
\end{figure}
\begin{figure}
\begin{tabular}{cc}
\includegraphics[height=15pc]{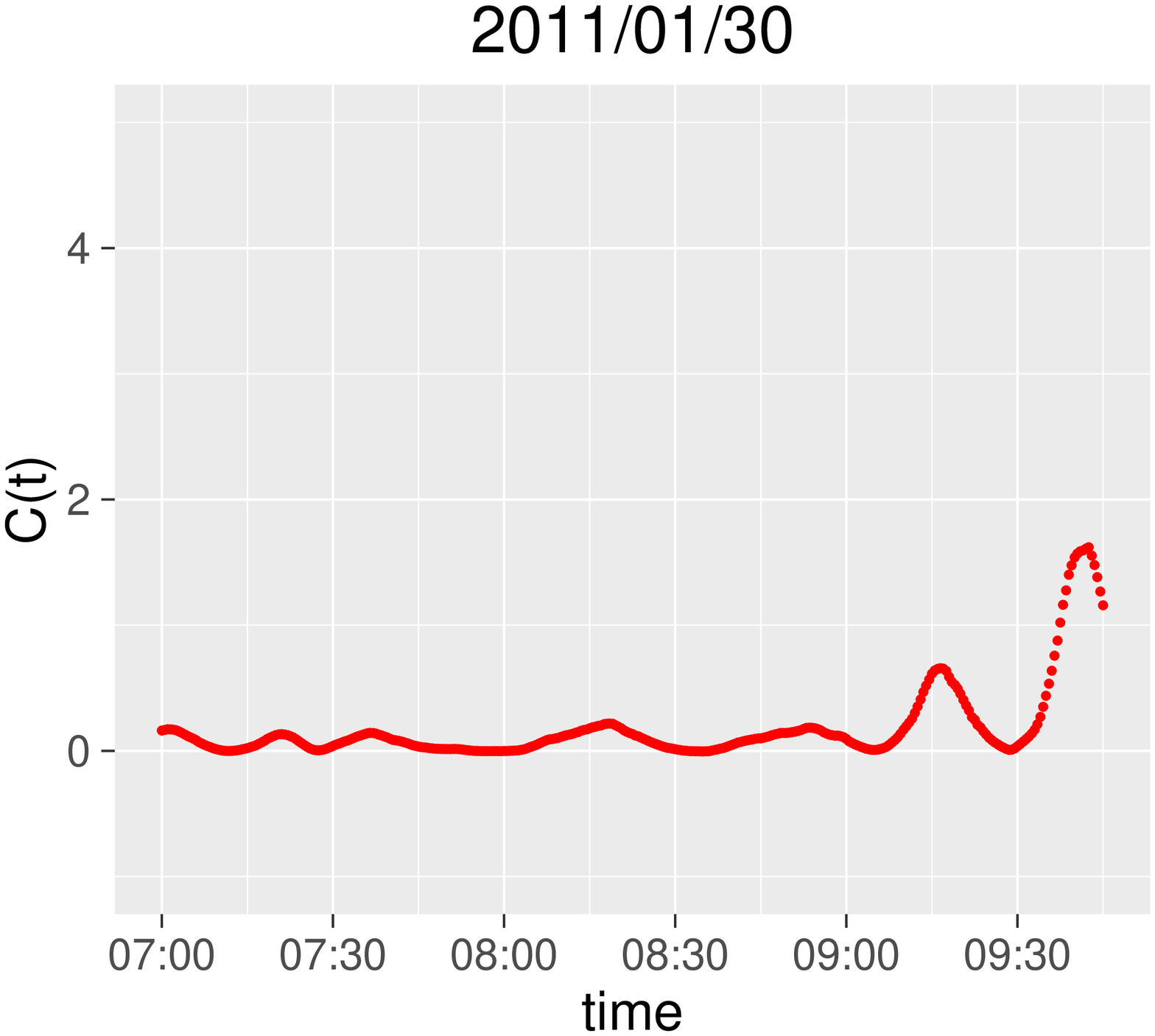} &
\includegraphics[height=15pc]{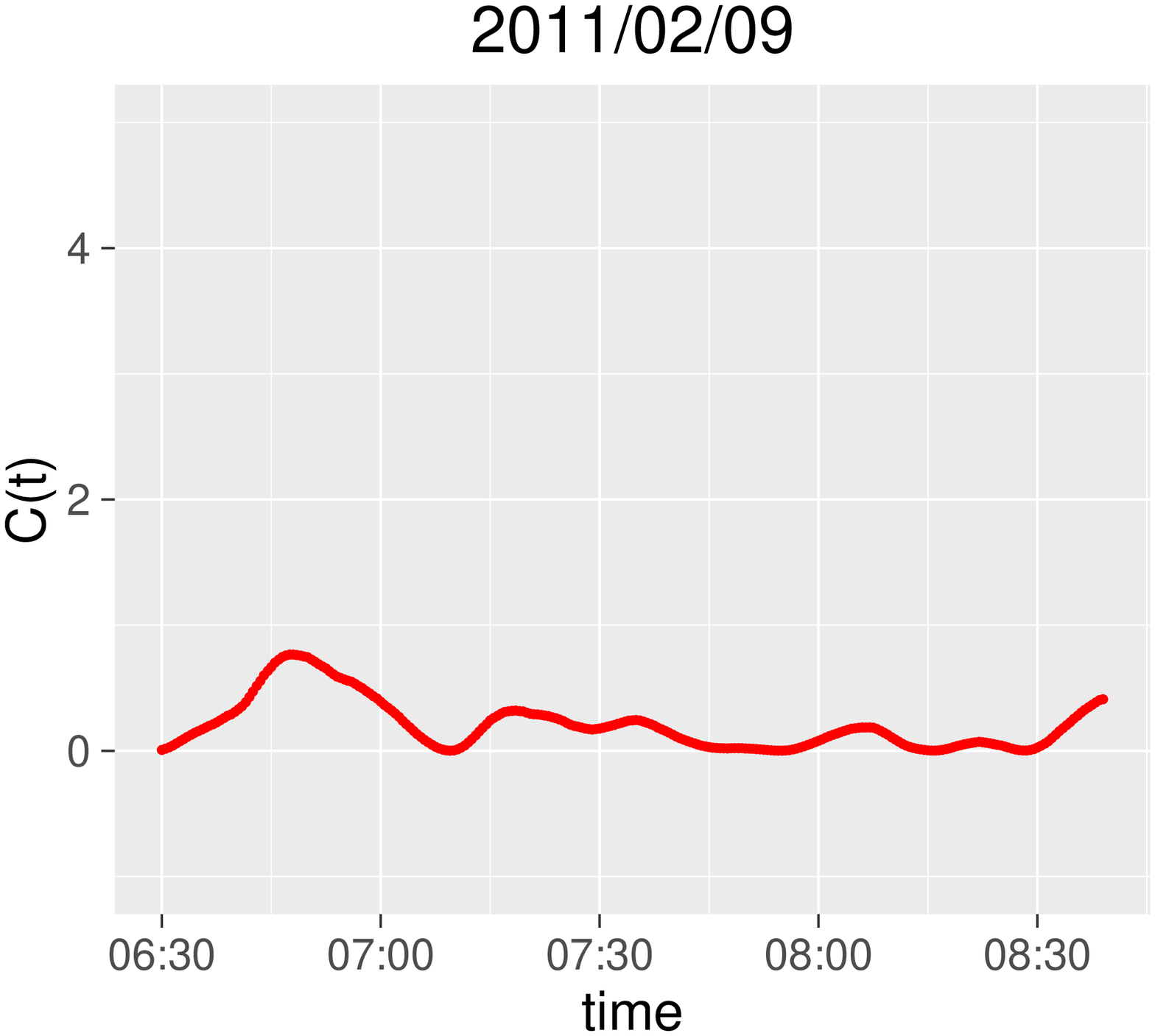}\\
\includegraphics[height=15pc]{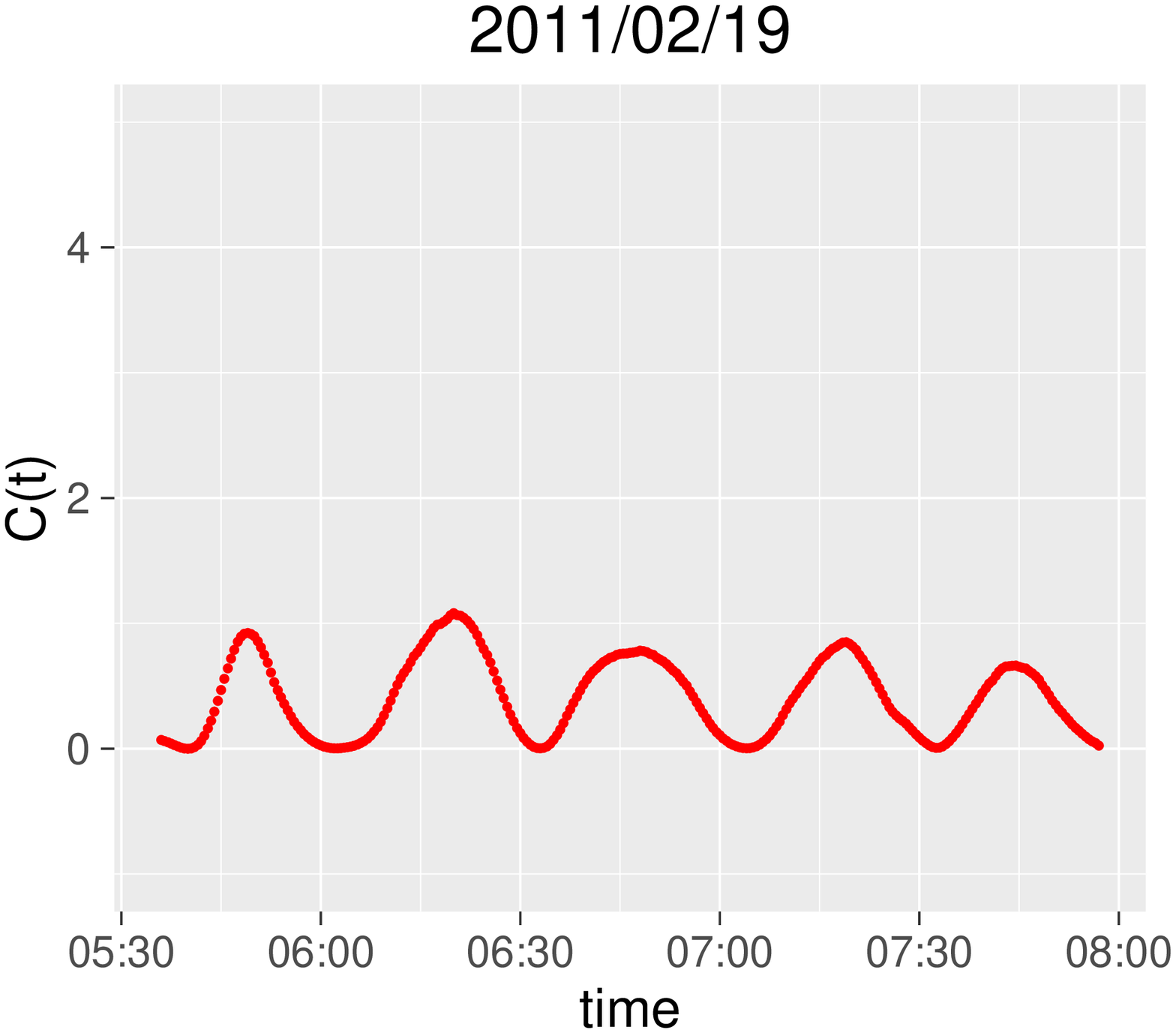} &
\includegraphics[height=15pc]{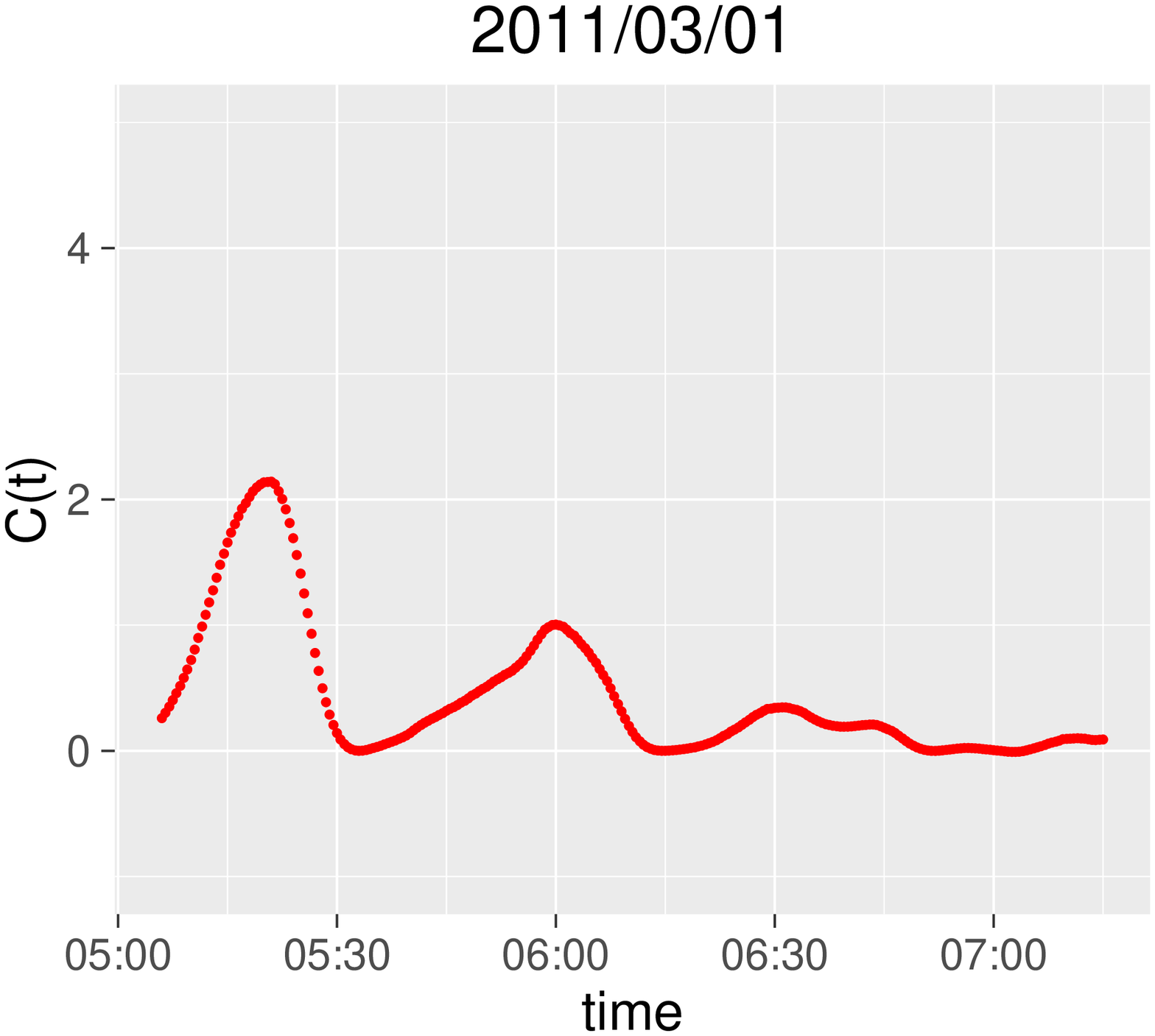}\\
\end{tabular}
\caption{The result of correlation analysis on non-earthquake days.  The vertical axis shows the correlation $C(T)$ and the horizontal one the time $t$ [UTC].  The day 40, 30, 20, 10 days before the earthquake, respectively. We used 5th polynomial functions as fitting curves.}
\label{kitaibaraki30-60p5}
\end{figure}
\begin{figure}
\begin{tabular}{cc}
\includegraphics[height=15pc]{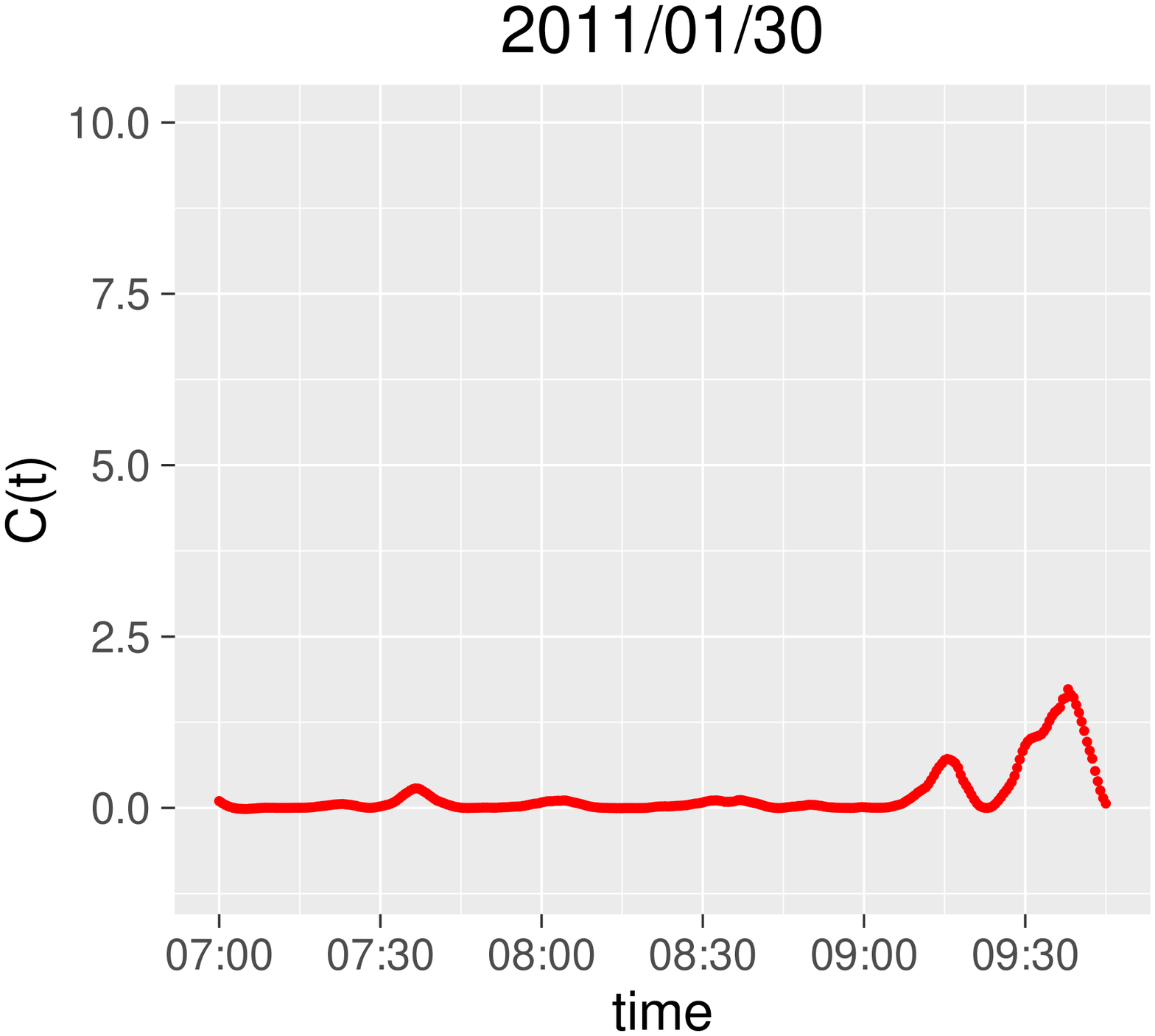} &
\includegraphics[height=15pc]{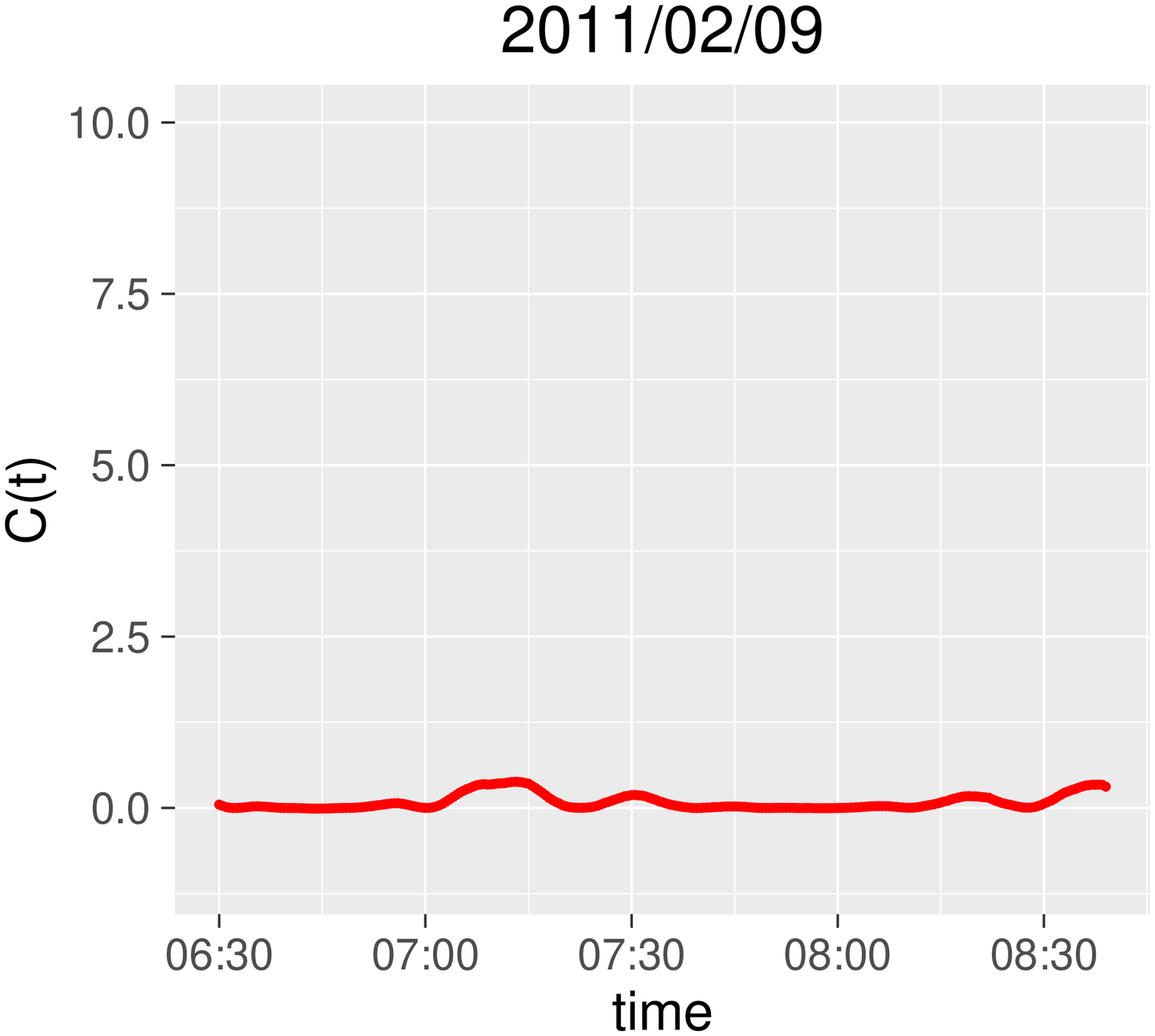}\\
\includegraphics[height=15pc]{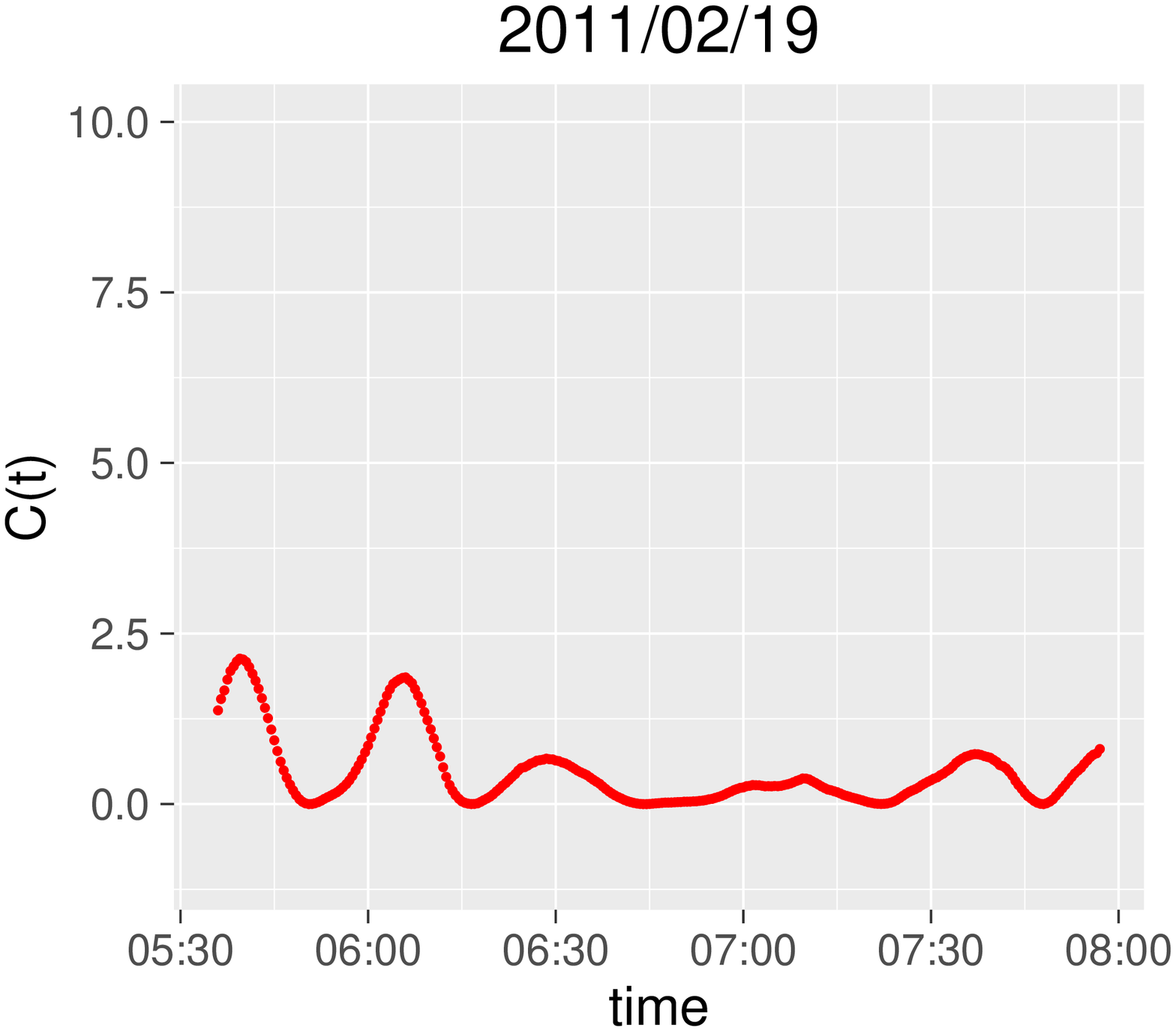} &
\includegraphics[height=15pc]{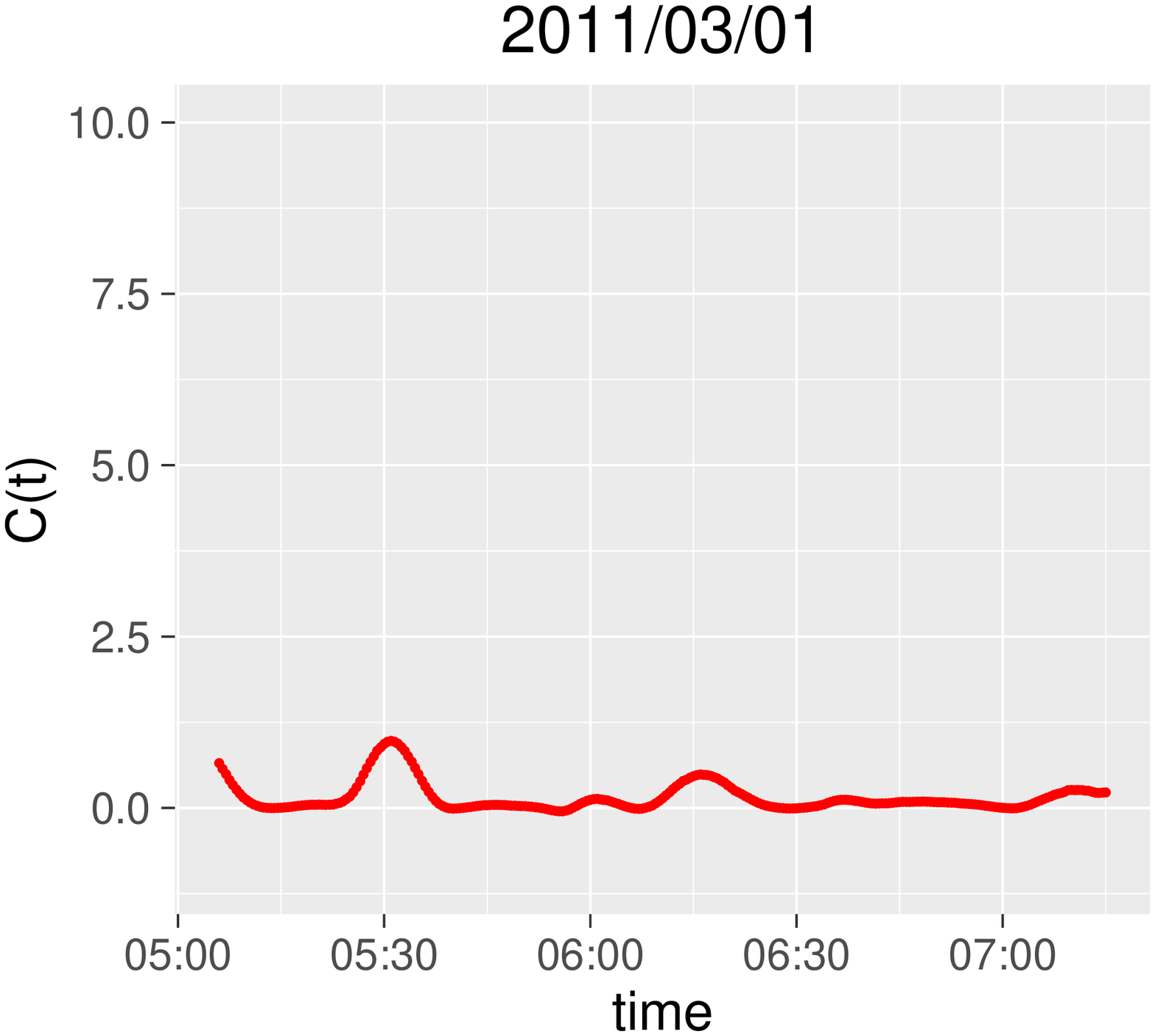}\\
\end{tabular}
\caption{The result of correlation analysis on non-earthquake days.  The day 40, 30, 20, 10 days before the earthquake, respectively. We used 3rd Fourier series as fitting curves.}
\label{kitaibaraki30-60f3}
\end{figure}
\begin{figure}
\begin{tabular}{cc}
\includegraphics[height=15pc]{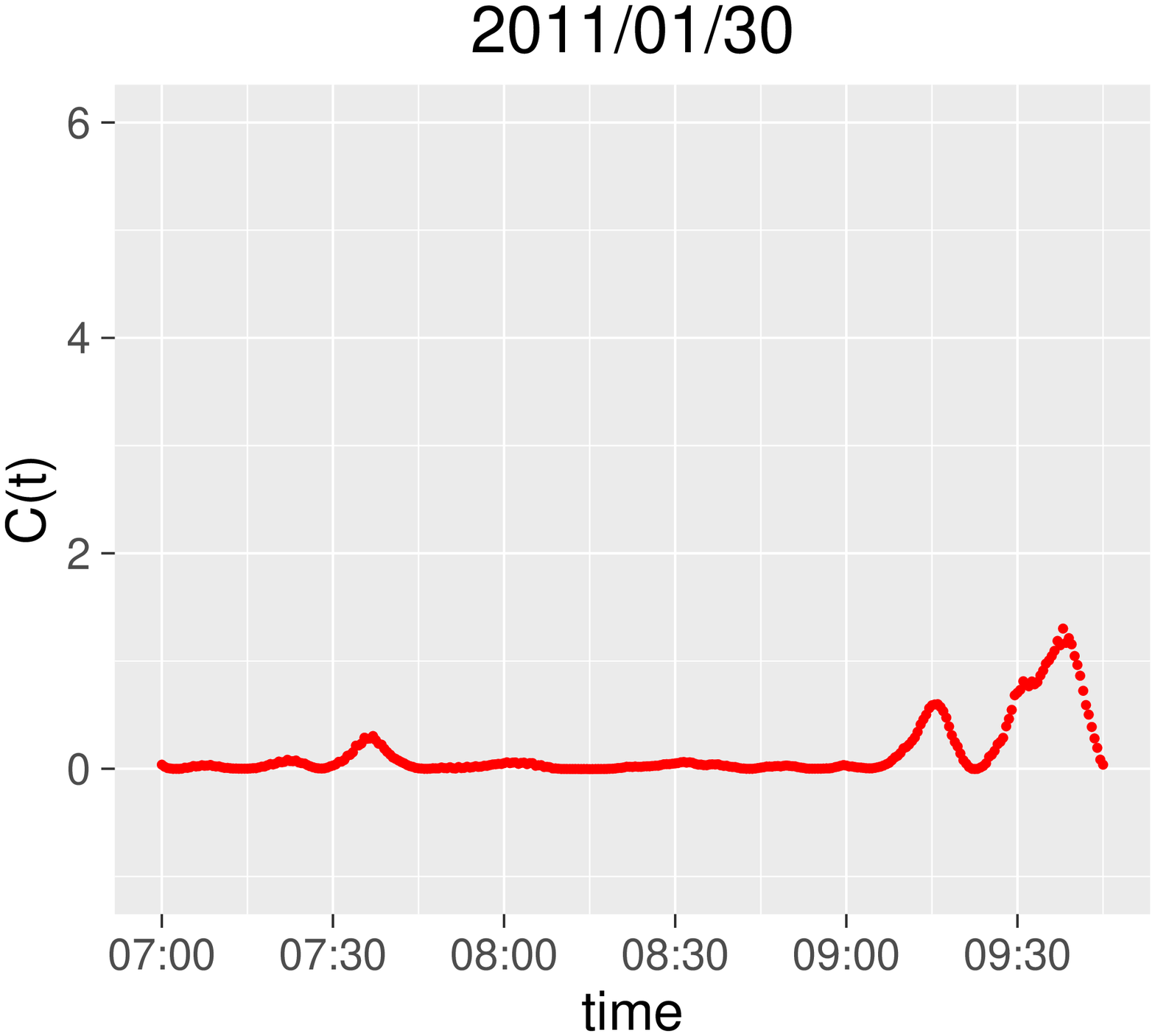} &
\includegraphics[height=15pc]{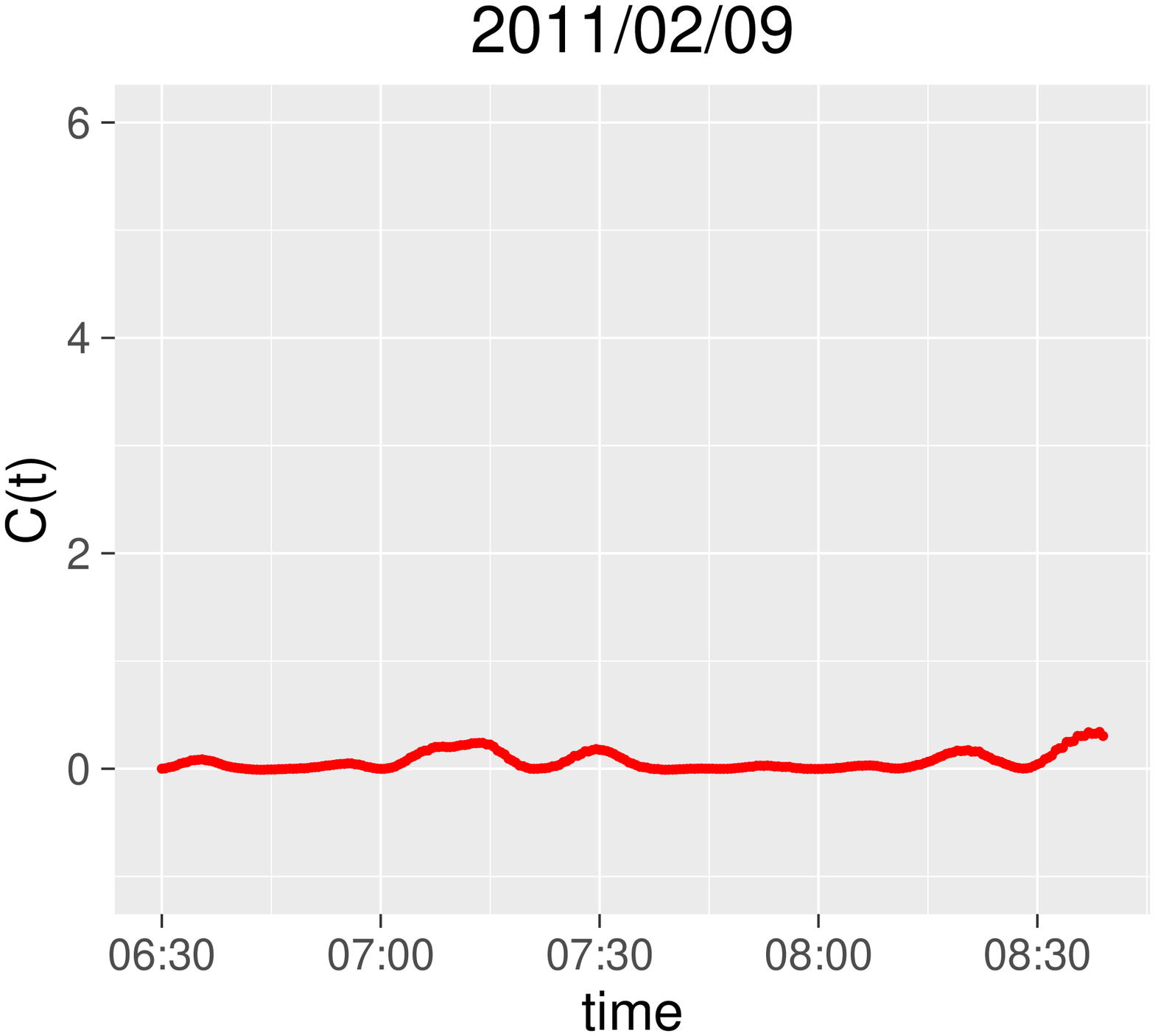}\\
\includegraphics[height=15pc]{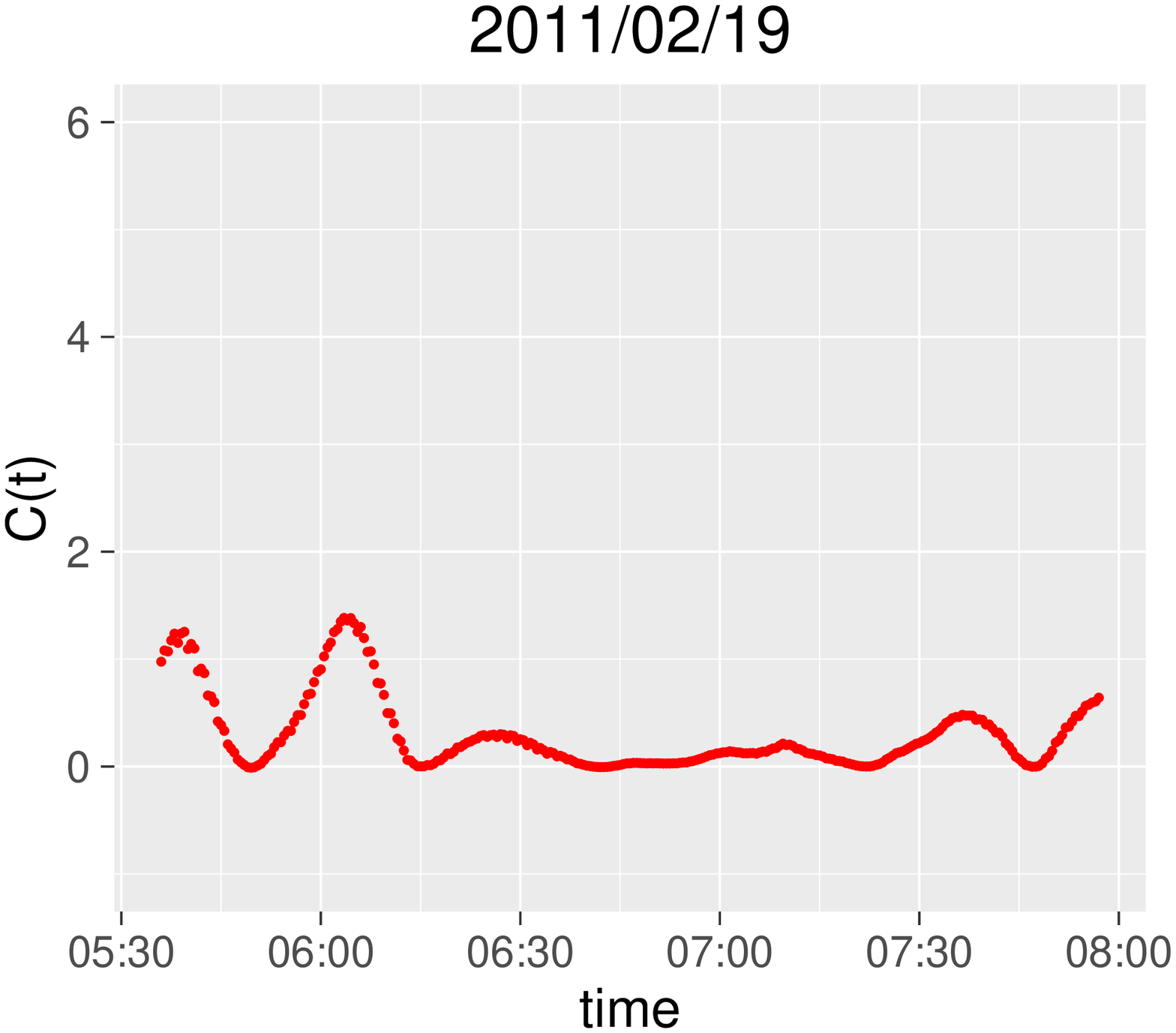} &
\includegraphics[height=15pc]{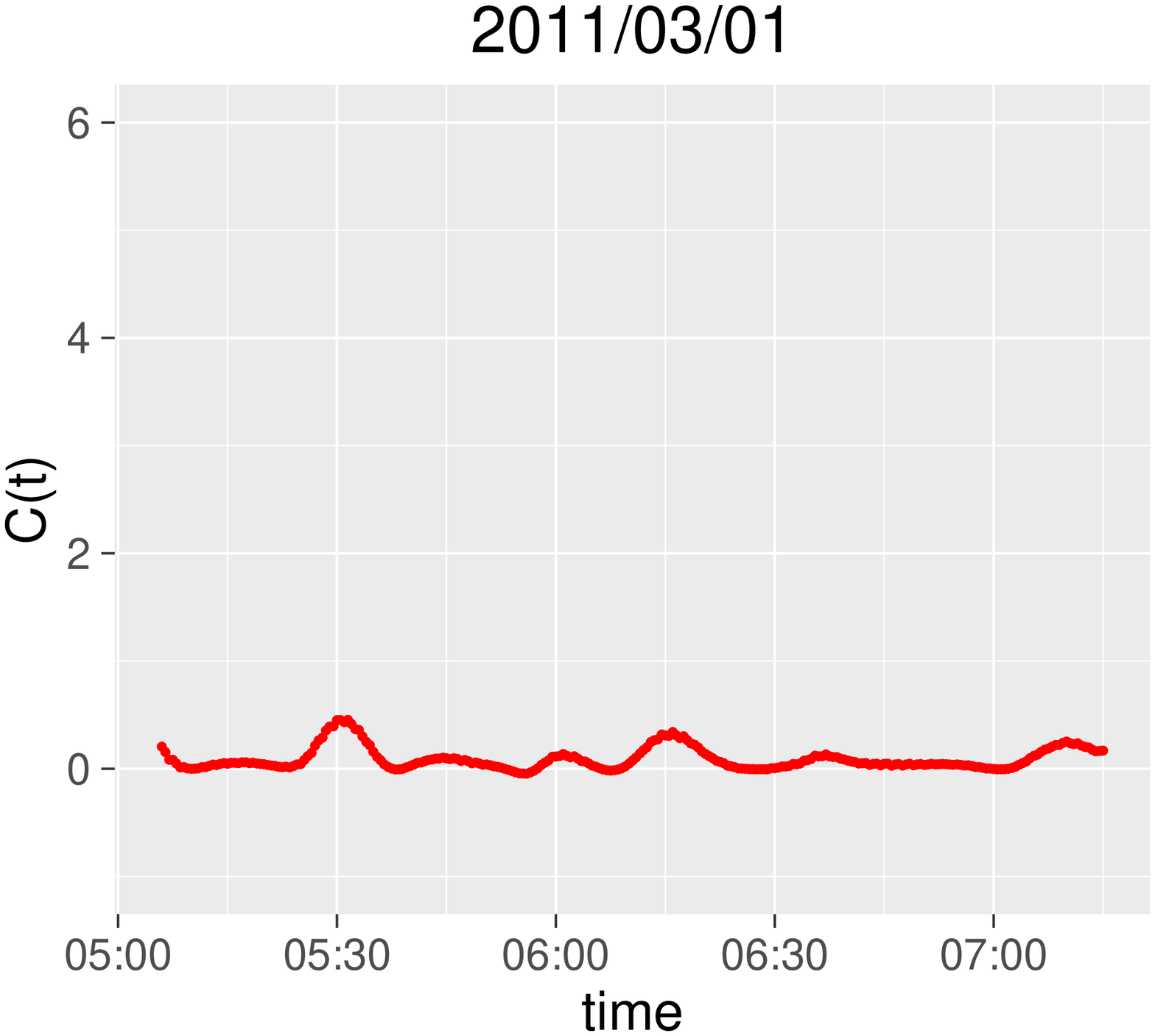}\\
\end{tabular}
\caption{The result of correlation analysis on non-earthquake days.  The day 40, 30, 20, 10 days before the earthquake, respectively. We used 7th Gaussian functions as fitting curves.}
\label{kitaibaraki30-60g7}
\end{figure}
\end{document}